\documentclass[11pt]{elsarticle}
\biboptions{sort&compress}
\usepackage{graphicx} % Required for inserting images
\usepackage{amssymb}
\usepackage{amsmath}
\usepackage{subcaption}
\usepackage[margin=1in]{geometry} 
\usepackage{lipsum}
\usepackage{makecell}
\usepackage{lineno}
\usepackage{lipsum}
\usepackage{algorithm}
\usepackage{algpseudocode}
\usepackage{tabularx} 
\usepackage{microtype}
\usepackage{comment}

\usepackage[]{siunitx}
\usepackage{xcolor}
\definecolor{lightblue}{rgb}{0.63, 0.74, 0.78}
\definecolor{seagreen}{rgb}{0.18, 0.42, 0.41}
\definecolor{orange}{rgb}{0.85, 0.55, 0.13}
\definecolor{silver}{rgb}{0.69, 0.67, 0.66}
\definecolor{rust}{rgb}{0.72, 0.26, 0.06}
\definecolor{purp}{RGB}{68, 14, 156}

\colorlet{lightrust}{rust!50!white}
\colorlet{lightorange}{orange!25!white}
\colorlet{lightlightblue}{lightblue}
\colorlet{lightsilver}{silver!30!white}
\colorlet{darkorange}{orange!75!black}
\colorlet{darksilver}{silver!65!black}
\colorlet{darklightblue}{lightblue!65!black}
\colorlet{darkrust}{rust!85!black}
\colorlet{darkseagreen}{seagreen!85!black}

\usepackage{hyperref}
\hypersetup{
  colorlinks=true,
}

\usepackage{cleveref}
\crefname{equation}{equation}{equations}
\crefname{figure}{figure}{figure}
\usepackage{booktabs}

\usepackage{fancyhdr}
\fancypagestyle{firstpage}{
    \fancyhf{}

}
\makeatletter
\def\ps@pprintTitle{%
    \let\@oddhead\@empty
    \let\@evenhead\@empty
    \let\@oddfoot\@empty
    \let\@evenfoot\@empty
    \def\@oddfoot{}%
    \def\@evenfoot{}%
}
\makeatother
\usepackage[table]{xcolor}
\begin{document}

\hypersetup{
  linkcolor=darkrust,
  citecolor=seagreen,
  urlcolor=darkrust,
  pdfauthor=author,
}

\thispagestyle{firstpage}
\begin{frontmatter}



\title{{\large\bf Hierarchical Bayesian constitutive model selection for high-strain-rate soft material characterization}}
%inertial microcavitation rheometry} 
% Article title


\author[Brown]{Victor Sanchez} %% Author name
\author[Brown]{Sawyer Remillard}
\author[UMich]{Bachir A. Abeid}
\author[UTexas1]{Lehu Bu}
%\author[Brown]{David Henann} %% Author name
\author[GT1,GT2,GT3]{Spencer~H.~Bryngelson}
\author[UTexas1,UTexas2]{Jin~Yang}
\author[UMich]{Jonathan B. Estrada}
\author[Brown]{Mauro Rodriguez Jr.} %% Author name

\affiliation[Brown]{organization={School of Engineering, Brown University},%Department and Organization
            % addressline={184 Hope Street}, 
            city={Providence},
            postcode={02912}, 
            state={RI},
            country={USA}}

\affiliation[UMich]{organization={Department of Mechanical Engineering, University of Michigan},%Department and Organization
            % addressline={1221 Beal Ave}, 
            city={Ann Arbor},
            postcode={48109}, 
            state={MI},
            country={USA}}            
\affiliation[GT1]{organization={School of Computational Science and Engineering, Georgia Institute of Technology},%Department and Organization
            % addressline={150 Bobby Dodd Way}, 
            city={Atlanta},
            postcode={30332}, 
            state={GA},
            country={USA}}
            
\affiliation[GT2]{organization={Daniel Guggenheim School of Aerospace Engineering, Georgia Institute of Technology},%Department and Organization
            % addressline={150 Bobby Dodd Way}, 
            city={Atlanta},
            postcode={30332}, 
            state={GA},
            country={USA}}
            
\affiliation[GT3]{organization={George W. Woodruff School of Mechanical Engineering, Georgia Institute of Technology},%Department and Organization
            % addressline={150 Bobby Dodd Way}, 
            city={Atlanta},
            postcode={30332}, 
            state={GA},
            country={USA}}
\affiliation[UTexas1]{organization={Materials Science and Engineering Program, Texas Materials Institute, The University of Texas at Austin},%Department and Organization
            % addressline={2617 Wichita St}, 
            city={Austin},
            postcode={78712}, 
            state={TX},
            country={USA}}
\affiliation[UTexas2]{organization={Department of Aerospace Engineering and Engineering Mechanics, The University of Texas at Austin},%Department and Organization
            % addressline={2617 Wichita St}, 
            city={Austin},
            postcode={78712}, 
            state={TX},
            country={USA}}

%% Abstract

\begin{abstract}
The high-fidelity characterization of soft, tissue-like materials under ultra-high-strain-rate conditions is critical in engineering and medicine.
Still, it remains challenging due to limited optical access, sensitivity to initial conditions, and experimental variability. 
Microcavitation techniques (e.g., laser-induced microcavitation) have emerged as a viable method for determining the mechanical properties of soft materials in the ultra-high-strain-rate regime (higher than $\SI{E3}{\per\second}$); however, they are limited by measurement noise and uncertainty in parameter estimation.
A hierarchical Bayesian model selection method is employed using the Inertial Microcavitation Rheometry (IMR) technique to address these limitations.
With this method, the parameter space of different constitutive models is explored to determine the most credible constitutive model that describes laser-induced microcavitation bubble oscillations in soft, viscoelastic, transparent hydrogels. 
The target data/evidence is computed using a weighted Gaussian likelihood with a hierarchical noise scale $\beta$, which enables the quantification of uncertainty in model plausibility.
Physically informed priors, including range-invariant, stress-based parameter priors, a model-redundancy prior, and a Bayesian Information Criterion motivated model prior, penalize complex models to enforce Occam's razor.
Using a precomputed grid of simulations, the probabilistic model selection process enables an initial guess for the \textit{Maximum A Posteriori} (MAP) material parameter values.
Synthetic tests recover the ground-truth models and expected parameters.
Using experimental data for gelatin, fibrin, polyacrylamide, and agarose, MAP simulations of credible models reproduce the data.
Moreover, a cross-institutional comparison of 10\% gelatin indicates consistent constitutive model selection.  
\end{abstract}


\begin{highlights}
\item A scalable Bayesian inference framework for high-strain-rate rheometry that performs data-driven model selection and quantifies uncertainty through a hierarchical noise formulation
\item Computationally validated a physics-based, reproducible soft material characterization approach in the high-strain-rate regime across independent laser-induced microcavitation datasets
\end{highlights}

%% Keywords

\end{frontmatter}
% \linenumbers

% {\Large\bf \textcolor{red}{please see this AI-generated review that has some good points: \url{https://chatgpt.com/share/68ebd4a0-1c7c-800a-a657-e4aafb7a4e4f}}}

\section{Introduction}

Soft biological and tissue-like materials experience extreme mechanical loading during events such as blast exposure, cavitation, and histotripsy, where deformation occurs on the microsecond timescale~\cite{Taylor2009,Marsh2021,Yeats2023,Maxwell2013}.
The combined rapid energy transfer and large deformations during these events place the material response in the high-strain-rate regime (\SIrange{1e3}{1e8}{\per\second})~\cite{Dougan2022,Estrada2017,Yang2020,haskell2023}.
Under such conditions, these materials exhibit nonlinear viscoelastic behavior, which is significantly different from their quasistatic response~\cite{Estrada2021,Eyal2016,Roberts2005,Finan2024,Barney2020,Lang2021,Rashid2012,Saraf2007,Franceschini2006,Siviour2016}.
% The characterization of these materials is particularly challenging in this regime. 
Accurately capturing and characterizing this behavior is important for applications ranging from impact biomechanics to focused ultrasound and soft matter engineering.
Yet, it remains difficult due to limited optical access, sensitivity to boundary conditions, and coupled thermal-diffusive effects that arise during rapid loading~\cite{Roberts2005,Finan2024,Yang2019,Yang2020,Johnsen2015,Barajas2017}.
Characterization is typically conducted in two parts: the determination of constitutive models and the estimation of parameters.
If a constitutive material model has been identified with a priori information in the parameter domain of interest, then this process is simplified to calibrating parameter values.

Conventional experimental approaches only offer partial insight into this extreme regime, but they are limited in several notable ways. 
Split-Hopkinson pressure bar or direct high-speed impact testing, for example, measures material response at strain rates of ($\SIrange{1e2}{1e4}{\per\second}$)~\cite{Chen2011,vanSligtenhorst2006,Zhang2011,Trexler2011,albrecht2013characterization,khatam2014dynamic} but is constrained by inertial effects, weak transmitted signals, and the assumption of stress equilibrium~\cite{Chen2016,Song2005,Seifert2023}. 
High-speed imaging with digital image correlation (DIC) or particle tracking has been used to capture full-field deformation of soft materials during impact or shock loading~\cite{Yang2021,Baldit2013,Ning2011,Yang2021thesis,Sugerman2023,Pan2009,mcghee2023high,yang2022serialtrack,buyukozturk2022high}.
However, DIC requires the sample surface to have dense speckle patterns and faces an inherent tradeoff between noise reduction and spatial resolution~\cite{Reu2018,Reu2022}. 
Particle tracking may introduce mismatch errors \cite{patel2018rapid}. 
Computational modeling techniques, such as finite element analysis (FEA) and multiscale constitutive modeling~\cite{Hrapko2008,Puglisi2016}, have been employed to simulate the rate-dependent and large-strain behavior of soft materials.
\citet{Hrapko2008} used FEA to model the dynamic behavior of brain tissue using various viscoelastic constitutive models and validated their simulations against shear experiments (\SIrange{0.1}{1}{\per\second}).
Their analysis revealed a strong sensitivity to model form and parameter values, posing a challenge to achieving consistent and physically reliable model selection in the quasistatic regime.
More recently, data-driven and machine learning approaches have been proposed to integrate experimental data into predictive frameworks~\cite{kirchdoerfer2016data,Bock2019}.
Although these methods can capture nonlinear and rate-dependent behavior, they often require extensive, high-quality datasets that are rarely available for soft materials~\cite{Upadhyay2024}. 
When data are sparse relative to the model complexity, overfitting can occur, while limited coverage of the relevant parameter or strain-rate space can lead to underfitting~\cite{Xu2023}.

Cavitation has proven to be an effective rheometry technique for probing the mechanical response of soft materials across a wide range of strain rates~\cite{Barney2020}.
Early work by \citet{Zimberlin2007} showed that needle-induced cavitation in hydrogels can be used to infer local stiffness.
Subsequent studies extended these approaches to biological tissues, including brain tissue, where cavitation dynamics have been used to extract viscoelastic and failure properties~\cite{Dougan2022,Dougan2024}.
More recent work has shown that inertial cavitation can provide meaningful rheological information in hydrogels at high strain rates. 
For example, \citet{Cohen2010} analyzed the dynamic expansion of a spherical cavity in compressible elastoplastic materials and identified conditions under which plastic shock waves appear. 
Related efforts in soft viscoelastic media by \citet{Warnez2015,Gaudron2015} incorporated rate-dependent dissipation and elastic nonlinearity into cavitation dynamics models, showing that viscosity can delay rapid cavity growth and attenuate shock-like responses.

Leveraging these prior insights, the Inertial Microcavitation Rheometry (IMR) technique was developed to enable the characterization of soft hydrogels across strain rates ranging from \SIrange{1e3}{1e8}{\per\second}~\cite{Estrada2017}.
Under varying pressures and material constraints~\cite{Plesset1977,Brennen1995,Johnsen2015,Abu2022,Yang2021b,Yang2019,Zhu2025,haskell2023}, the radial dynamics of microcavitation are modeled using modified Keller--Miksis equations.
In IMR, materials are characterized by comparing laser-induced microcavitation experimental observations to Keller--Miksis-type forward numerical calculations for simple viscoelastic constitutive models using a least-squares fitting approach~\cite{Estrada2017}. 
Experimental radius-time data are typically extracted near the rebound peaks, when the bubble wall velocity is low and optical measurements are most reliable.
High-speed imaging at frame rates greater than one million frames per second captures most of the bubble dynamics.
However, rapid motion near collapse may fall outside the temporal resolution or show motion blur.

The IMR technique considers multiple viscoelastic constitutive material models~\cite{Yang2020,Yang2021b,Yang2019,abeid160experimental} to characterize hydrogels, e.g., gelatin, agarose, fibrin, and polyacrylamide. 
\citet{Spratt2021} extended the technique using ensemble-based data assimilation methods, obtaining similar mechanical properties for the same materials as \citet{Estrada2017} while reducing computational cost.
Their results suggested that achieving a best fit between simulation and experiment may require a viscosity that evolves during the bubble collapse, potentially reflecting damage or structural softening effects not captured by conventional constitutive models~\cite{Spratt2021}.
\citet{Zhu2025} developed a parsimonious IMR technique that simplifies soft material characterization by focusing on the initial bubble collapse time rather than fitting the entire radius--time history.
Both advances reduced computational complexity and calibration efforts; however, the methods are unable to account for experimental noise and model-form uncertainty.

Bayesian inference provides a probabilistic approach for extending the IMR technique to model selection and parameter estimation.
Unlike deterministic least-squares methods that yield a single best-fit solution, Bayesian inference can quantify the plausibility of competing constitutive models by evaluating model performances through their relative probabilities~\cite{Sandeep2015,Oden2013,Chiachio2015,Babuska2016} and then identify the associated parameters~\cite{Muto2008,Elmukashfi2022,Fitzenz2007,Rosic2013,Sarkar2012}.
This probabilistic formulation directly evaluates model accuracy and quantifies uncertainty without relying on extensive prior information~\cite{Gelman2021,Beck1998,Katafygiotis1998}.
Additionally, Bayesian inference naturally incorporates model-form priors, experimental uncertainty, and measurement noise across multiple datasets~\cite{Beck1998,Rappel2019,Most2010}, making it well-suited for reproducible and uncertainty-aware characterization of soft materials.

Recent studies have shown the growing impact of Bayesian inference in rheological problems.
\citet{chu25,chu252} applied Bayesian optimal experimental design to bubble dynamics and soft material characterization, showing that information-theoretic criteria can accelerate data acquisition and improve inference efficiency. 
Similarly, \citet{Freund2015} used Bayesian model selection to determine the optimal number of modes in a multi-modal Maxwell model, which describes the dynamic shear moduli of a synthetic polymer.
Their results showed that Bayesian inference can reliably discriminate between competing viscoelastic models with high-dimensional parameter spaces and noisy experimental data.
In this context, Bayesian inference favors constitutive models that strike a balance between goodness of fit and physical interpretability, rather than empirical or highly flexible formulations with many parameters that risk overfitting.
While effective in selecting a model, their method was unable to quantify uncertainty. 

To this end, this work aims to implement Bayesian inference in conjunction with the IMR technique to perform model selection across competing constitutive models under high-strain-rate conditions, while accounting for uncertainty quantification.
The work is organized as follows.
In \cref{sec:theory}, we present the governing equations for the bubble dynamics used to generate forward simulations across a parameter space of constitutive model parameters, the experimental setup for laser-induced cavitation (LIC), the material characterization procedure, and the Bayesian inference IMR implementation.
The results of the Bayesian IMR characterization for soft materials and discussion of their implications are presented in \cref{sec:results,sec:discussion}, respectively.
We summarize the capabilities of the Bayesian IMR technique and outlook in \cref{sec:conclusions}.

\section{Theory and methods\label{sec:theory}}

\subsection{Governing equations\label{subsec:physics}}

\begin{figure}
    \centering
    \includegraphics[width=0.6\linewidth]{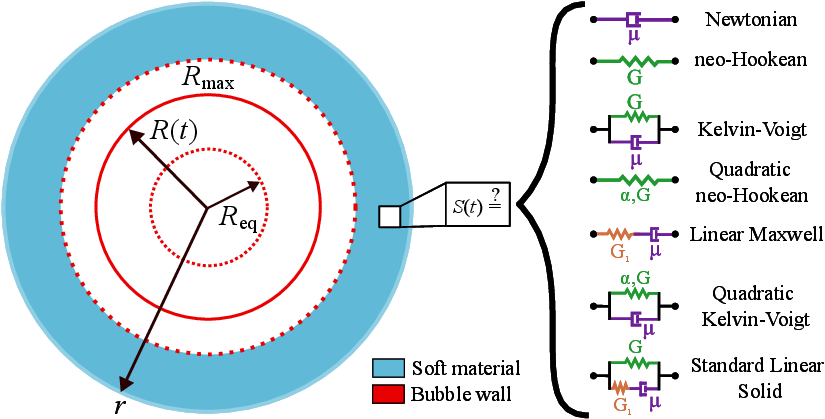}
    \caption{A spherical bubble is nucleated in a soft material and grows to a maximum radius before collapsing and oscillating until it reaches mechanical equilibrium. Identification of the stress integral formulation enables the characterization of the material.}
    \label{fig:problem_setup}
\end{figure}

We consider a spherical bubble that is nucleated in a soft material by a laser-induced cavitation event.
As shown in \cref{fig:problem_setup}, the bubble expands to a maximum radius and subsequently collapses and oscillates toward a mechanical equilibrium.
To model these dynamics, the IMR forward simulation code base is used to perform numerical simulations of the governing equations that describe bubble oscillations. 
The equations consist of an ordinary differential equation (ODE) for the bubble's radial dynamics, which is coupled with partial differential equations (PDEs) for the energy balance inside the bubble, and a PDE for mass transfer between the bubble's water vapor and non-condensible gas~\cite{Warnez2015,Barajas2017}. 
A numerical resolution of 100 points inside the bubble was used for the forward simulations.
In the present study, stress evolves through explicit functions or ODEs, as defined in \cref{tab:models}~\cite{Estrada2017,Gaudron2015,Zhu2025}. 
For brevity, only the radial equation for the bubble is stated here.
The bubble's radial dynamics are modeled using the Keller–Miksis equation~\cite{Keller1980,Prosperetti1986a}:
\begin{equation}
    \left(1-\frac{\dot{R}}{c}\right)R\ddot{R} 
    + \frac{3}{2} \left(1-\frac{\dot{R}}{3c}\right)\dot{R}^2 
    = \frac{1}{\rho}\left(1+\frac{\dot{R}}{c}\right)
    \left(p_{\text{b}} - p_{\infty} - \frac{2\gamma}{R} + S \right)
    + \frac{1}{\rho}\frac{R}{c}\dot{\overline{\left(p_{\text{b}} - \frac{2\gamma}{R} + S\right)}},
    \label{eq:keller-miksis}
\end{equation}
where the dot denotes a time derivative, $R$ is the bubble radius, $\dot{R}$ the bubble wall velocity, $c$ the speed of sound in the surrounding material, $\rho$ its density, $S$ the material viscoelastic contribution of the stress integral, $\gamma$ the surface tension between the material and the vapor and gas filled bubble, and $p_{\text{b}}$ and $p_\infty$ the bubble and far-field pressures, respectively. 
$S$ is defined as 
\begin{equation}
S(t) = 2\int_{R}^{\infty} 
    \frac{\tau_{rr} - \tau_{{\theta}{\theta}}}{r}\,\mathrm{d}r,
    \label{eq:stress_integral}
\end{equation}
where $\tau_{rr}$ and $\tau_{{\theta}{\theta}}$ are the respective radial and hoop components of the deviatoric Cauchy stress predicted by the chosen constitutive model, and $r$ the radial coordinate in the surrounding medium.
% The far-field boundary condition enforces $s_{rr}=s_{{\theta}{\theta}}=0$ as $r\to\infty$.

For comparison across materials, material models, and experimental conditions, it is beneficial to non-dimensionalize the governing equations.
We present the governing equations in non-dimensional form.
We choose the bubble radius at its maximum size, $L_c=R_{\text{max}}$, as the characteristic length, the far-field pressure and material density to define the characteristic velocity, $v_{\text{c}}=\sqrt{p_\infty/\rho}$, and the characteristic time, $t_{\text{c}}=L_c/v_{\text{c}}=R_{\text{max}}\sqrt{\rho/p_\infty}$~\cite{Estrada2017,Brennen1995,Jomni2009}.
The non-dimensionalized variables are
\begin{equation*}
    R^*=\frac{R}{R_{\max}},\quad
    t^*=\frac{t}{t_{\text{c}}},\quad
    c^*=\frac{c}{v_{\text{c}}},\quad
    p_{\text{b}}^*=\frac{p}{p_\infty},\quad 
    \text{and} \quad
    S^*=\frac{S}{p_\infty}.
\end{equation*}
Thus, the non-dimensional Keller--Miksis equation~\cite{Estrada2017},
\begin{align}
    % \label{eq:nondimKMeq}
    \left(1-\frac{\dot{R}^*}{c^*}\right) R^* \ddot{R}^* + &\frac{3}{2}\left(1-\frac{\dot{R}^*}{3 c^*}\right) \dot{R}^{*2} =\\& \left(1+\frac{\dot{R}^*}{c^*}\right)\left(p^*_{\text{b}}-1-\frac{1}{\text{We}R^*}+S^*\right)\notag +\frac{R^*}{c^*}\dot{\overline{\left(p^*_{\text{b}}-\frac{1}{\text{We}R^*}+S^*\right)}}.
\end{align}
The explicit forms of $S^*$ for each constitutive model are tabulated in \cref{tab:models}, expressed in terms of the wall stretch ratio $\lambda=R^*/R_{\text{eq}}^*$, where $R_{\text{eq}}^*$ denotes the non-dimensional equilibrium bubble radius.
The dimensionless parameters are the Weber number, Reynolds number, Cauchy number, and Deborah number defined as,
\begin{equation*}
    \text{We} =\frac{\rho v_{\text{c}}^2R_{\text{max}}}{\gamma}, \quad \text{Re} =\frac{\rho v_{\text{c}}R_{\text{max}}}{\mu}, \quad \text{Ca} = \frac{p_\infty}{G}, \quad \text{De} =\frac{\lambda_1}{t_\text{c}},
\end{equation*}
respectively, where $\mu$ is the surrounding material shear viscosity, $G$ the elastic shear modulus, $\lambda_1 = \mu/G_1$ is the relaxation time.
The constitutive model parameter ranges are tabulated in \cref{tab:modelParams}.
The strain-stiffening parameter, $\alpha$, is also non-dimensional.
We fix the total number of forward simulations per model to $4096$ and construct a Cartesian grid with $n^d=4096$ points, where $d$ is the number of model parameters and $n=4096^{1/d}$. 
Consequently, the per-axis resolution decreases with increasing dimensionality (e.g., $n=4096$ for $d=1$, $n=64$ for $d=2$, $n=16$ for $d=3$), ensuring comparable total computational effort across models of different dimensions.

\begin{table}[t!]
    \centering
    \renewcommand{\arraystretch}{2}
    \caption{Constitutive material models and corresponding stress integral.
    }
    \begin{tabular}{l l l}
        \toprule
        \textbf{Tag}&\textbf{Constitutive model} & \textbf{Stress integral} $\displaystyle \left(\lambda= R^*/R_{\text{eq}}^*\right)$
        \\ 
        \midrule
         Newt &Newtonian fluid&
         $\displaystyle S^*_{\text{v}} =-\frac{4}{\text{Re}}\frac{\dot{R}^*}{R^*}$
         \\
         
         NH&neo-Hookean solid &
         $\displaystyle S^*_{\text{NH}}=\frac{1}{2\text{Ca}}\left[4\lambda^{-1}+\lambda^{-4}-5\right]$
         \\
         
         KV&Kelvin-Voigt solid&
         $\displaystyle S^*_{\text{KV}}=S^*_{\text{v}}+S^*_{\text{NH}}$
         \\

         qNH&Quadratic neo-Hookean solid &
         % \makecell{
         % $S^*_{\text{qNH}}=(3\alpha-1)S^*_{\text{NH}}+}$ \\
         % \frac{2\alpha}{\text{Ca} \left\[ \frac{27}{40} +\frac{1}{8}\lambda^{8}+\frac{1}{5}\lambda^{5}+\lambda^{2}-\frac{2}{\lambda} \right\]$
         % }
        \makecell[l]{
        $\displaystyle S^*_{\text{qNH}} = (1-3\alpha) S^*_{\text{NH}} + 
        $ \\[1ex]
        \qquad$\displaystyle \frac{2\alpha}{\text{Ca}}  \left[ \frac{27}{40} + \frac{1}{8}\lambda^{-8} + \frac{1}{5}\lambda^{-5} + \lambda^{-2} - 2\lambda \right]$} \\
         LM&Linear Maxwell fluid &$\text{De}\dot{S}^*_{\text{M}}+S^*_{\text{M}}=S^*_{\text{v}}$\\

         qKV&Quadratic Kelvin-Voigt solid &
        $\displaystyle S^*_{\text{qKV}}=S^*_{\text{qNH}}+S^*_{\text{v}}$\\
         
         SLS&Standard Linear Solid& $\displaystyle S^*_{\text{SLS}}=S^*_{\text{M}}+S^*_{\text{NH}}$\\
         \bottomrule
    \end{tabular}
    \label{tab:models}
\end{table}

\begin{table}[t!]
\centering
    \caption{Simulation material parameters for soft-tissues in the high-strain rates and their non-dimensional counterparts~\cite{Rashid2012,Estrada2017,Spratt2021}.}
\renewcommand{\arraystretch}{1.1}
    \begin{tabular}{ l c c c } 
        \toprule
        \textbf{Parameter} &\textbf{Notation}&\textbf{Minimum value}&\textbf{Maximum value}\\
        \midrule
        Viscosity [\SI{}{\pascal\second}] & $\mu$ & $10^{-4}$ & $1$\\ 
        Reynolds number & Re & 1.6& $\SI{36E3}{}$\\
        Shear modulus [\SI{}{\pascal}]  & $G$ & $10^2$ & $10^5$ \\ 
        Cauchy number & Ca &  0.2 & $\SI{E3}{}$ \\
        Relaxation time [\SI{}{\second}] & $\lambda_1$ & $10^{-7}$ & $10^{-3}$ \\ 
        Deborah number & De & \SI{3E-3}{} & 65 \\
        Strain stiffening  & $\alpha$& $10^{-3}$& 10  \\
        \bottomrule
    \end{tabular}
    \label{tab:modelParams}
\end{table}

\subsection{Experiments}

\begin{figure}[t!]
    \centering
    \includegraphics[width=0.45\linewidth]{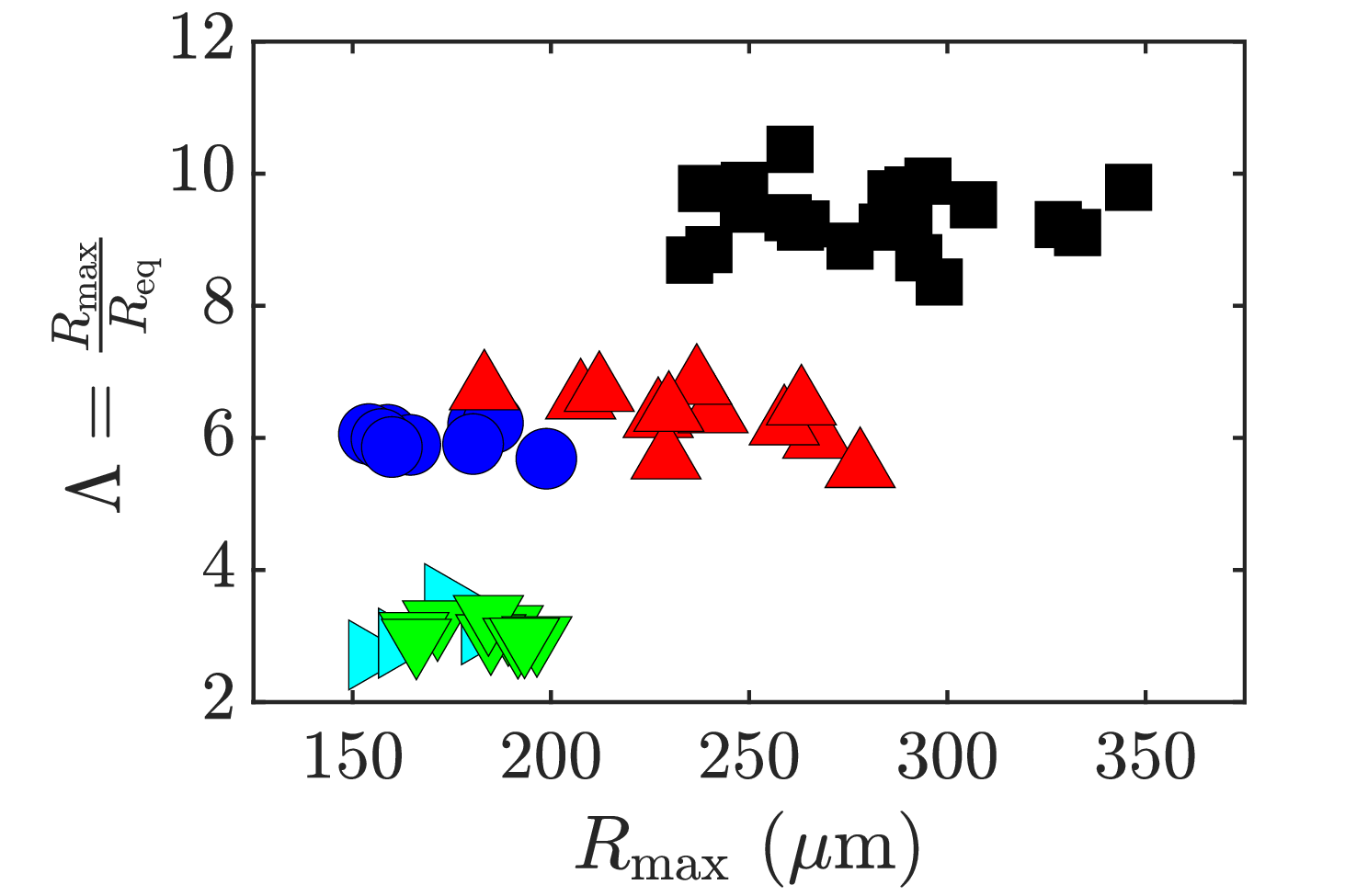}
    \caption{Experimental maximum stretch ratio v. maximum radius.
    UM1: blue circle, UM2: black square, UM3: red up triangle, UT1: cyan right triangle, UT2: green down triangle.
    }
    \label{fig:lambda_vs_Rmax}
\end{figure}

LIC experiments in soft materials were conducted and captured using a high-speed camera~\cite{Yang2020,Maxwell2013,Wilson2019,haskell2023}. 
The University of Michigan (UM) and the University of Texas at Austin (UT) LIC experimental setups, material preparation protocols, and datasets for the hydrogels are  in~\cite{abeid160experimental} and~\cite{Yang2021}, respectively.
The materials and their corresponding polymer concentrations are tabulated in \cref{tab:materials}, and a comparison of the maximum bubble radius $R_{\text{max}}$ with the maximum stretch ratio $\Lambda = \text{max}(\lambda)$ is shown in \cref{fig:lambda_vs_Rmax}. 
Each dataset consists of multiple experimental trials for one material.
% Trials were conducted for gelatin, fibrin, and an acrylamide/bisacrylamide combination (polyacrylamide, PAAm) from UM, and for gelatin and agarose from UT. 
A subset of UM hydrogel concentrations were used in the present work. 
Gelatin data of the same concentration from both institutions were analyzed to cross-validate the constitutive model selection.
% LIC experiments in gelatin were also conducted by UT and the samples were prepared as follows. 
% Gelatin Type A powder (611995000; Thermo Scientific) was mixed with ultrapure deionized  water. 
% The resulting solution was heated to \SI{50}{\celsius} in a water bath until it became homogeneous.
% Then, it was transferred to petri dishes, where it was cooled to room temperature to form solid gels.
The experimental setup generates and records a single LIC event in transparent hydrogels at ultra-high-speed frame rates ($\geq$\SI{1e6}{} fps).
The image processing is done using an in-house bubble edge detection MATLAB code~\cite{InCA2025}.

% Datasets consisting of multiple experimental trials were obtained for gelatin, fibrin, and an acrylamide/bisacrylamide combination (polyacrylamide, PAAm) from the University of Michigan (UM), and for gelatin and agarose from the University of Texas at Austin (UT). 

\begin{figure}[t!]
    \centering
    \includegraphics[width=0.32\textwidth]{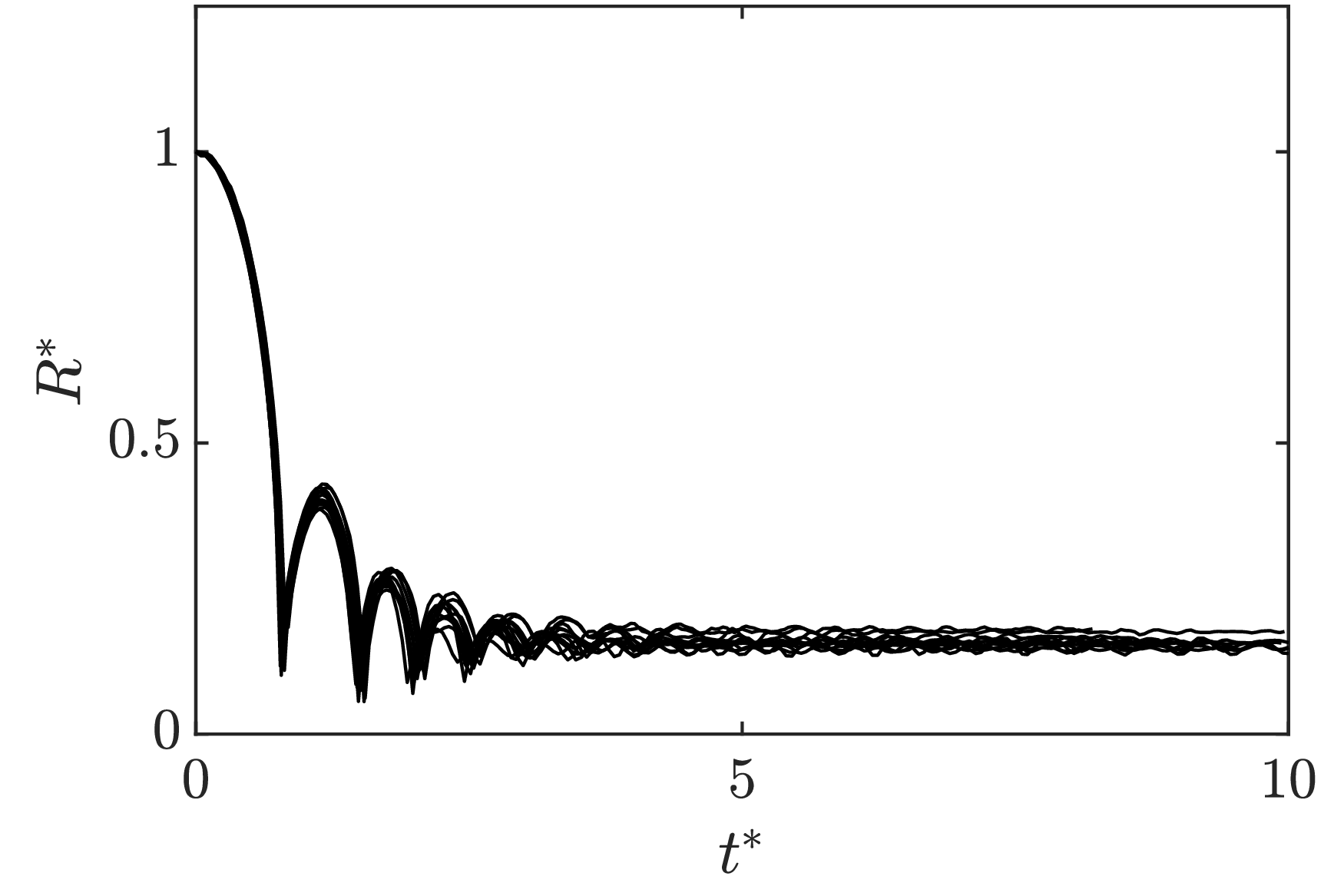}
    \includegraphics[width=0.32\textwidth]{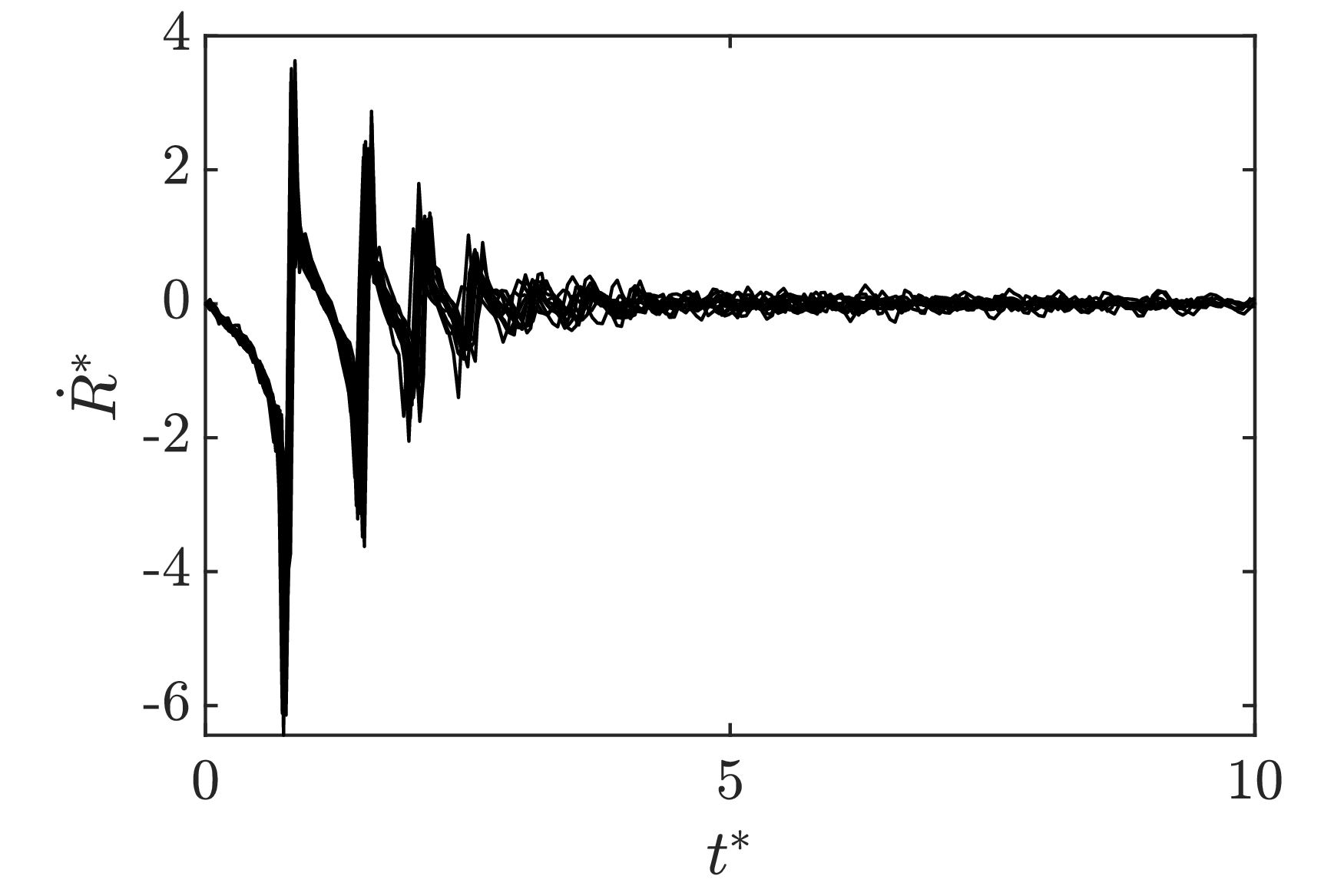}
    \includegraphics[width=0.32\textwidth]{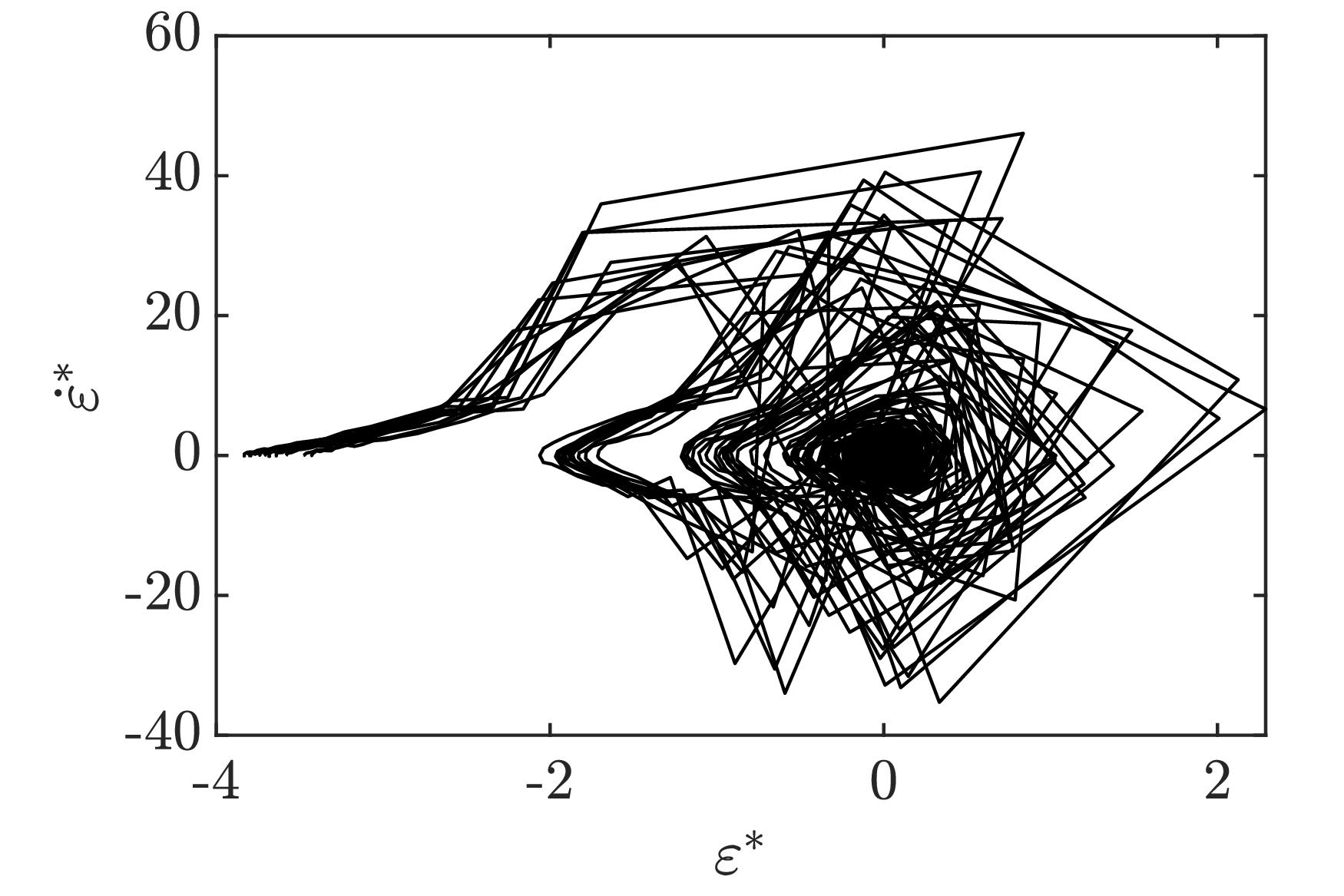}
    \caption{
    Bubble radius (left), velocity (middle), and magnitude of strain rates at the bubble wall (right) histories for nine sets of gelatin LIC experimental data.
    }
    \label{fig:expdata}
\end{figure}

In \cref{fig:expdata}, the left panel shows normalized bubble radius histories for PAAm, where differences between trials primarily reflect variations in the applied laser energy, with added contributions from measurement noise, slight material heterogeneity, and minor differences in viscous response due to varying $R_{\text{max}}$.
The middle panel shows the corresponding bubble wall velocities, computed from the non-dimensional experimental $R^*$ data using a Padé approximation.
The bubble wall radius and velocities are used to compute the non-dimensional strain rate and strain via the logarithmic Hencky strain defined in~\cref{eq:filter}.
The right panel shows the phase space between the non-dimensional strain rate and strain, which are considered for filtering out low-information data points in \cref{subsec:Pre-process}. 
Although the normalized LIC curves collapse into a consistent temporal window, variations in the collapse and rebound amplitudes across trials show the combined effects of experimental noise and material variability. 
These observations motivate the use of probabilistic model comparison, as deterministic fits do not adequately represent the observed data spread.

\begin{table}[t!]
\centering
    \caption{Institution, materials, and concentrations of the laser-induced microcavitation experiments.}
\renewcommand{\arraystretch}{1.1}
    \begin{tabular}{ l l c | l l c } 
        \toprule
        \textbf{Tag}& \textbf{Material} &\textbf{Concentration [wt\%]} & \textbf{Tag}& \textbf{Material} &\textbf{Concentration [wt\%]}\\
        \midrule
        UM1   & Gelatin & 10.0  & UT1   & Gelatin & 10.0 \\ 
        UM2   & Fibrin & 0.2 & UT2   & Agarose & 5.0   \\ 
        UM3   & PAAm & 8.0/0.26 \\%& UT3   & PEGDA & \mrdz{number?} \\ 
        \bottomrule
    \end{tabular}
    \label{tab:materials}
\end{table}

\subsubsection{Pre-processing}\label{subsec:Pre-process}

For a given soft material, there is a dataset of $J$ experimental trials (radius-time curves) indexed as $j = 1, \dots, J$.
Each trial is non-dimensionalized by its own peak radius, $R_{\text{max},j}$, and characteristic time, $t_{\text{c},j}$.
After pre-processing, each trials provides a set of non-dimensional time stamps $\{t^*_{j,k}\}_{k\in\mathcal{K}_j}$ and corresponding bubble radius and bubble wall velocity measurements $\{R^*_{j,k}, \dot{R}^*_{j,k} \}_{k \in \mathcal{K}_j}$, where $R^*_{j,k}\in(0,1]$ and $\mathcal{K}_j$ denotes the index set of valid time-step indices for trial $j$. 
In some trials, the experimental signal extends far beyond the collapse events of interest, with small oscillations about equilibrium that contribute disproportionately to the Gaussian likelihood.
To focus the Bayesian analysis on informative regions, $\mathcal{K}_j$ is restricted using a combination of strain-based filters.
We define the radial normal component of both the Hencky strain with respect to the per-trial equilibrium radius, $R^*_{\text{eq},j}$, and the dimensionless strain rate as 
\begin{equation}
    \label{eq:filter}
    \varepsilon^*_{j,k}=\frac{1}{2}\ln\left(\left(\frac{R^*_{j,k}}{R^*_{\text{eq},j}}\right)^{-4}\right),
    \quad
    \dot\varepsilon^*_{j,k}=-2\frac{\dot{R}^*_{j,k}}{R^*_{j,k}}.
\end{equation}
We set threshold values at
\begin{equation}
    \varepsilon^*_{\text{th},j}=0.1\max_{k}|\varepsilon^*_{j,k}|,
    \quad
    \dot\varepsilon^*_{\text{th},j}=10^5 ~t_{\text{c},j},
\end{equation}
where the strain threshold corresponds to the onset of finite-strain behavior in soft hydrogels, beyond which small-strain assumptions no longer hold~\cite{Gandin2021}, and the strain rate threshold corresponds to the high-strain-rate regime.
Time-steps are retained if they lie outside an elliptical gate in the $(\varepsilon^*_j,\dot\varepsilon^*_j)$ plane,
\begin{equation}
    \left(\frac{\varepsilon^*_{j,k}}{\varepsilon^*_{\text{th},j}}\right)^2+\left(\frac{\dot\varepsilon^*_{j,k}}{\dot\varepsilon^*_{\text{th},j}}\right)^2\geq 1,
    \label{threshold}
\end{equation}
which removes near-equilibrium points characterized by low strain and low strain rate, while preserving the collapse-rebound regions of interest.

Each simulation is generated using the mean values extracted from the experiments, 
\begin{equation}
    \overline{R}_{\text{max}} = \frac{1}{J}\sum_{j=1}^{J} R_{\text{max},j},
    \qquad
    \overline{t}_{\text{c}} = \frac{1}{J}\sum_{j=1}^{J} t_{\text{c},j}.
\end{equation}
For a given model $M_i$ and associated parameter vector $\boldsymbol{\theta}_i$, where $i \in [1, N_M]$ and $N_M$ is the number of models, the forward simulation produces a discrete time history $\{t_{\text{s},m}^*, R^*_{\text{s},m}, \dot{R}^*_{\text{s},m}\}$, where the subscript $\text{s}$ indicates simulation data and $m=1,\dots,N_t$ indexes the simulation time steps with maximum $N_t$ time steps. 
The simulation data are then interpolated to the experimental time stamps $\{t^*_{j,k}\}$ to obtain the simulated radius and velocity at the proper time steps, denoted $R^*_{\text{s}}(t^*_{j,k}; M_i, \boldsymbol{\theta}_i)$ and $\dot{R}^*_{\text{s}}(t^*_{j,k}; M_i, \boldsymbol{\theta}_i)$. 
Residuals between the experimental observations and the simulated predictions are defined as
\begin{subequations}
    \begin{equation}
        r_{R^*,jk}(M_i,\boldsymbol{\theta}_i) = R^*_{j,k} - R^*_{\text{s},jk}(M_i,\boldsymbol{\theta}_i),
    \end{equation}
    \begin{equation}
        r_{\dot R^*,jk}(M_i,\boldsymbol{\theta}_i) = \dot R^*_{j,k} - \dot R^*_{\text{s},jk}(M_i,\boldsymbol{\theta}_i),
    \end{equation}    
\end{subequations}
where $\boldsymbol{\theta}_i = \{\theta_1, \dots, \theta_n\}$ is the set of $n$ model parameters for model $M_i$. 

\subsection{Characterization}\label{sec:characterize}
For a single material and fixed laser input energy, repeated LIC events reveal variability between experimental realizations.
A 95\% confidence interval computed across multiple trials provides statistical bounds on the expected variation at each time step.
However, the underlying noiseless response may not be fully captured within the available experimental datasets. 
While the pre-processing step filters out low-information regions to limit their influence, residual uncertainty remains in both the retained and discarded portions of the data. 
The Bayesian inference approach introduced in \cref{subsec:BIMR-approach} incorporates this uncertainty through the likelihood formulation, which explicitly accounts for observational noise and potential model–data mismatch. 
Since the forward simulations solve deterministic governing equations, the modeling noise is assumed to be zero. 
However, the probabilistic formulation still enables consistent treatment of measurement variability across trials.

\subsubsection{Bayes' theorem for model selection}\label{subsec:BIMR-approach}

A Bayesian approach is used to select a constitutive model and quantify the agreement between the IMR forward simulation results and the LIC experimental data. 
Bayes' theorem for model selection is~\cite{Sivia2006,Gregory2005,Held2020}
\begin{equation}
    P(M_i|D,I)=\frac{P(M_i|I)P(D|M_i,I)}{P(D|I)},
    \label{eq:BayesThm}
\end{equation}
where $P(M_i|D,I)$ is the posterior probability of a model, $P(M_i|I)$ the model prior, $P(D|M_i,I)$ the observational likelihood, and $P(D|I)$ the marginal likelihood for $i$-th model, $M_i$, $D$ the given data, and $I$ the system constraints.
% We also introduce a normalized observational likelihood comparison term,
% \begin{equation}
%     \mathcal{L}_i=\frac{P(D|M_i,I)}{\sum_i P(D|M_i,I)}.
% \end{equation}
For simplicity and consistency with the literature, the constraint term is omitted from the notation, though it is understood to be present.
Additionally, an intrinsic Occam's Razor penalizes overly complex models, yielding the simplest model that maximizes the observational likelihood, where complexity refers to the number of model parameters~\cite{Blanchard2017}.
For IMR characterization, $D$ is the experimental and numerical simulation bubble radius-time and bubble wall velocity-time data, i.e., $R^*(t^*)$ and $\dot{R}^*(t^*)$.

\subsubsection{Hierarchical noise scaling and likelihood}

Baseline heteroscedastic variance estimates are computed per time step, denoted by $\sigma^2_{0,R^*}(t^*_{j,k})$ and $\sigma^2_{0,\dot{R}^*}(t^*_{j,k})$.
To model the time‐varying reliability of experimental measurements, we apply a logistic weight that depends on the instantaneous strain rate,
\begin{equation}
    a_{j,k}=\frac{1}{1+\exp\left[-\kappa\left(\frac{\dot\varepsilon^*_{\mathrm{th},j}-|\dot\varepsilon^*_{j,k}|}{\dot\varepsilon^*_{\mathrm{th},j}}\right)\right]},
\end{equation}
where $\kappa$ controls the steepness of the transition around the threshold strain rate $\dot{\varepsilon}^*_{\mathrm{th},j}$.
$\kappa$ was set to unity to have a gradual transition that is neither abrupt nor overly diffuse, allowing the weighting to change on a physically meaningful scale comparable to typical variations in measured strain rate. 
Tests with larger $\kappa$ values produced sharper transitions that offered no improvement in model discrimination and occasionally reduced numerical robustness.
The logistic function is then mapped to a bounded weight \(w_{j,k}\) via
\begin{equation}
    w_{j,k} = m + (1-m)a_{j,k},\qquad
    m\in(0,1], \qquad \kappa>0,\label{lkhd_weight}
\end{equation}
where $m$ sets the minimum allowable weight, with the maximum being unity.
The value of $m$ is set to $0.1$, limiting down-weighting for data points in low-activity regions by at most a factor of ten relative to the highest-weighted points.
This small floor prevents near-zero weights from inflating the variance.
In this formulation, measurement uncertainty in high-speed optical imaging increases primarily with bubble wall velocity, and thus with strain rate.
Slow rebound or near-equilibrium dynamics yield minimal blur and tracking error, while rapid bubble collapses result in reduced resolution due to motion and finite camera exposure.
Therefore, including strain amplitude in the weighting would penalize optically reliable, slower‐moving segments and misrepresent the true measurement noise. 
In contrast, strain‐rate‐based weighting isolates the physically dominant source of heteroscedasticity without redundantly penalizing regions already excluded by the elliptical gate defined above.
The scaled variances are
\begin{subequations}
    \begin{equation}
        \sigma^2_{R^*}(t^*_{j,k};\beta) = 
            \frac{\beta^2\,\sigma_{0,R^*}^2(t^*_{j,k})}{w_{j,k}}, \quad
        \sigma^2_{\dot{R}^*}(t^*_{j,k};\beta) =
            \frac{\beta^2\,\sigma_{0,\dot{R}^*}^2(t^*_{j,k})}{w_{j,k}},    
    \end{equation}
\end{subequations}
where $\beta$ is a positive noise scale such that $\beta>0$ and inflates the per-time experimental variance.
Furthermore, $\beta$ serves as a diagnostic of data quality and as a safeguard against overfitting.
Models that require artificially large $\beta$ values to reconcile simulations with data are automatically downweighted in the plausibility calculation.
Thus, as shown in \cref{eq:betamarginal}, high posterior probabilities are reserved for models that explain the data within a realistic noise margin. 
This interpretability is particularly valuable when multiple viscoelastic formulations achieve similar log-likelihoods.
The combined consideration of plausibility and $\beta$ distributions distinguishes between models that genuinely represent the dynamics and those that rely on the inflated noise scaling. 

A simulation/model-discrepancy variance, $\sigma^2_{\text{s},R^*}(t^*_{j,k};M_i,\boldsymbol{\theta}_i)$ and $\sigma^2_{\text{s},\dot{R}^*}(t^*_{j,k};M_i,\boldsymbol{\theta}_i)$, can be included to account for uncertainty in stochastic or approximate forward models. 
The total variances are then
\begin{subequations}
    \begin{equation}
        \mathrm{Var}_{R^*}(t^*_{j,k};M_i,\boldsymbol{\theta}_i,\beta) = \sigma_{R^*}^2(t^*_{j,k};\beta) + \sigma^2_{s,R^*}(t^*_{j,k};M_i,\boldsymbol{\theta}_i),
    \end{equation}
    \begin{equation}
        \mathrm{Var}_{\dot{R}^*}(t^*_{j,k};M_i,\boldsymbol{\theta}_i,\beta) = \sigma_{\dot{R}^*}^2(t^*_{j,k};\beta) + \sigma^2_{s,\dot{R}^*}(t^*_{j,k};M_i,\boldsymbol{\theta}_i).
    \end{equation}
\end{subequations}
Here, the simulation/model-discrepancy variance terms are set to zero by default because the IMR forward simulations are deterministic.

To obtain $P(D|M_i)$, the parameters and noise scale are marginalized.
When modeling physical systems, Gaussian noise is commonly introduced to account for the inherent experimental variability and uncertainty, maximizing statistical entropy~\cite{Jaynes2003,Sivia2006}. 
A Gaussian observational likelihood is adopted,
\begin{equation}
\begin{split}
P(D|M_i,\boldsymbol{\theta}_i,\beta)
&=\prod_{j}\prod_{k\in\mathcal{K}_j}
\frac{1}{\sqrt{2\pi\,\mathrm{Var}_{R^*}(t^*_{j,k}\,;M_i,\boldsymbol{\theta}_i,\beta)}}
\exp\left[-\frac{r_{R^*,jk}(M_i,\boldsymbol{\theta}_i)^2}{2\,\mathrm{Var}_{R^*}(t^*_{j,k}\,;M_i,\boldsymbol{\theta}_i,\beta)}\right]
\\
&\quad\times
\frac{1}{\sqrt{2\pi\,\mathrm{Var}_{\dot{R}^*}(t^*_{j,k}\,;M_i,\boldsymbol{\theta}_i,\beta)}}
\exp\left[-\frac{r_{\dot{R}^*,jk}(M_i,\boldsymbol{\theta}_i)^2}{2\,\mathrm{Var}_{\dot{R}^*}(t^*_{j,k}\,;M_i,\boldsymbol{\theta}_i,\beta)}\right].
\label{eq:obsLklhd}
    \end{split}
\end{equation}

\subsubsection{Priors}

We employ a hierarchical prior structure that acts at three coupled levels: the noise-scale prior $P(\beta)$, the parameter prior $P(\boldsymbol{\theta}_i|M_i)$ within each constitutive model, and the model prior $P(M_i)$ across the set of candidate models. 
In principle, one could adopt a uniform prior over a considered space, $P(\cdot) = 1/V$, where $V$ is either the total number of elements or the volume in the space, depending on if it is discrete or continuous.
Such a prior corresponds to the state of maximum statistical entropy and is therefore the least biased choice in the absence of prior information~\cite{Sivia2006,Jaynes2003,Gelman2021,Box2011}. 
However, when models differ in dimensionality or physical formulation, the uniform assumption may not sufficiently penalize model complexity and assign similar plausibility to parameter regions that are physically redundant or poorly constrained. 
Therefore, we construct priors to penalize model complexity.
The noise prior constrains the level of variance inflation permitted when reconciling simulations with experimental data.
The parameter space prior ensures consistent weighting across physically meaningful and unique regions of the parameter space.
The model prior then imposes an Occam penalty to prevent overparameterization. 

To marginalize out $\beta$ from \cref{eq:obsLklhd}, we compute
\begin{equation}
    P(D|M_i,\boldsymbol{\theta}_i)=\int_0^\infty P(D|M_i,\boldsymbol{\theta}_i,\beta)P(\beta)d\beta,
    \label{eq:betamarginal}
\end{equation}
where 
\begin{equation}
    P(\beta)=\frac{2}{\pi}\frac{1}{1+\beta^2}, \quad \beta>0,
\end{equation}
is the Half-Cauchy distribution.
A Half-Cauchy prior for $\beta$ is a weakly informative, heavy-tailed distribution that regularizes small scales while allowing large values when supported by the data.
Thus, it is a robust default for variance components in hierarchical Bayesian models~\cite{gelman2006}.

Because the integral in \cref{eq:betamarginal} has no closed form antiderivative for the likelihood function used here, it is evaluated numerically by discretizing the domain of $\beta$ into $B$ grid points $\{\beta_{b}\}_{b=1}^{B}$,
\begin{equation}
    P(D|M_i,\boldsymbol{\theta}_i)\approx\sum_{b=1}^B q_b P(D|M_i,\boldsymbol{\theta}_i,\beta_{b}),
\end{equation}
where the normalized quadrature weights are defined as
\begin{equation}
    q_b \propto P(\beta_{b})\,\Delta\beta,
    \qquad
    \sum_{b=1}^{B} q_b = 1,
\end{equation}
and $\Delta\beta$ is the uniform grid spacing.
Although the Half--Cauchy prior is defined on a semi-infinite domain, its heavy-tailed form decays as $P(\beta)\sim \beta^{-2}$ for large $\beta$, meaning that most of its probability mass is concentrated near $\beta\lesssim10$.
Thus, we truncate the integration domain to the finite interval $[\beta_{\min},\beta_{\max}]=[0.05,10]$, which captures more than $99.9\%$ of the total prior mass with negligible truncation error. 
Within this finite range, the quadrature is performed on a uniform grid using the trapezoidal rule, with each grid point being the midpoint of a local interval between its neighbors. 
This symmetric construction preserves second-order accuracy and ensures that both interior and boundary points contribute proportionally to the prior-weighted integral. 

To marginalize out $\boldsymbol{\theta}_i$, we compute
\begin{equation}
    P(D|M_i)=\int\cdots\int P(D|M_i,\boldsymbol{\theta}_i)P(\boldsymbol{\theta}_i|M_i)d\boldsymbol{\theta}_i.
    \label{eq:thetamarginal}
\end{equation}
Similar to \cref{eq:betamarginal}, $P(D| M_i,\boldsymbol{\theta}_i)$ has no closed-form antiderivative in $\boldsymbol{\theta}_i$ for the likelihoods considered here. 
Thus, we evaluate \cref{eq:thetamarginal} numerically on a  Cartesian grid $\{\boldsymbol{\theta}_i^{(g)}\}_{g=1}^{N_g}$:
\begin{equation}
    P(D| M_i)\approx\sum_{g=1}^{N_g} w_g P\left(D| M_i,\boldsymbol{\theta}_i^{(g)}\right),
    \label{eq:theta_quadrature}
\end{equation}
where $g$ indexes each unique grid point and $N_g$ is the total number of parameter combinations.
The quadrature weights $w_g$ represent the normalized prior mass assigned to the cell at $\boldsymbol{\theta}_i^{(g)}$:
\begin{equation}\label{eq:priorWeight}
    w_g \propto P\left(\boldsymbol{\theta}_i^{(g)}|M_i\right)\Delta V_g,
    \qquad
    \sum_{g=1}^{N_g} w_g = 1,
\end{equation}
where $\Delta V_g$ is the (hyper-)volume of the grid cell.

\begin{figure}
    \centering
    \includegraphics[width=0.5\linewidth]{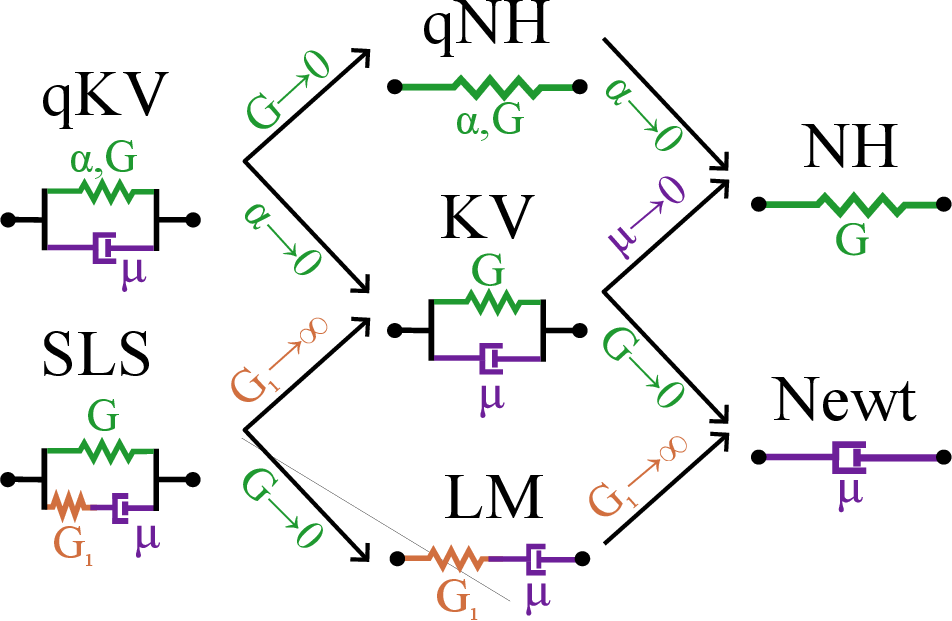}
    \caption{Schematic diagram of the hierarchical relationship among the constitutive models. 
    Arrows indicate limiting cases obtained by setting parameters to zero, thereby reducing the model's dimensionality.
    The first column of models corresponds to three parameters, the second column has two parameters, and the third column has one parameter.}
    \label{fig:model_hierarchy}
\end{figure}

We now define the prior for the parameter space to compute~\cref{eq:priorWeight}.
For each constitutive model $M_i$, the parameter grid $\boldsymbol{\theta}_i$ is mapped to a normalized space using logarithmic coordinates.
This normalization makes the parameter representation range-invariant with respect to the choice of parameter bounds.
The model's stress integral response is evaluated along the gated experimental segments defined in \cref{threshold}, denoted by $S^*_i(t^*_{j,k};\boldsymbol{\theta}_i)$.
These stress integrals are used to determine whether additional parameters of $M_i$ generate distinct stress integral histories relative to lower-dimensional models in the hierarchy shown in \cref{fig:model_hierarchy}.

To quantify this distinction, we consider a complex parent model and a reduced simpler model such that $M_\mathcal{S}\subseteq M_\mathcal{C}$ where $\boldsymbol{\theta}_\mathcal{S}\subseteq\boldsymbol{\theta}_\mathcal{C}$ denotes the corresponding parameter subsets.
Before comparing the stresses, we align their amplitudes using a scaling factor 
\begin{equation}
    c^* =
\frac{\sum_{j,k} w_{j,k}
S^*_{\mathcal{C}}(t^*_{j,k};{\boldsymbol{\theta}}_{\mathcal{C}})
S^*_{\mathcal{S}}(t^*_{j,k};{\boldsymbol{\theta}}_{\mathcal{S}})}
     {\sum_{j,k} w_{j,k}
      S^*_{\mathcal{S}}(t^*_{j,k};{\boldsymbol{\theta}}_{\mathcal{S}})^2},
\end{equation}
where $w_{j,k}$ are the gating weights from \cref{lkhd_weight}.
This scaling removes differences that arise purely from the overall stress integral magnitude between two models.

The relative stress mismatch between two models is then defined by 
\begin{equation}
    \Phi_{\mathcal{C}\rightarrow\mathcal{S}}({\boldsymbol{\theta}}_\mathcal{C})=\frac{||S^*_\mathcal{C}(t^*_{j,k};{\boldsymbol{\theta}}_\mathcal{C})-c^*S^*_\mathcal{S}(t^*_{j,k};{\boldsymbol{\theta}}_\mathcal{S})||_w}{||S^*_\mathcal{C}(t^*_{j,k};{\boldsymbol{\theta}}_\mathcal{C})||_w},
    \label{eq:PhiCtoS}
\end{equation}
where $||\cdot||_w$ denotes the weighted $\ell^2$ norm induced by $\{w_{j,k}\}$.
Small values of $\Phi_{\mathcal{C}\rightarrow\mathcal{S}}$ indicate that the additional parameters in $M_\mathcal{C}$ do not introduce stress integral histories that are distinguishable from those of the simpler model $M_\mathcal{S}$.

To convert this mismatch into a prior penalty, we define a dimensionless redundancy factor
\begin{equation}
    w_{\mathcal{C}\rightarrow\mathcal{S}}({\boldsymbol{\theta}}_{\mathcal{C}})=\frac{\Phi_{\mathcal{C}\rightarrow\mathcal{S}}      ({\boldsymbol{\theta}}_{\mathcal{C}})^2}{\Phi_{\mathcal{C}\rightarrow\mathcal{S}}({\boldsymbol{\theta}}_{\mathcal{C}})^2+ \tau_{\mathcal{S}}^2},
    \label{eq:wCtoS}
\end{equation}
where $\tau_\mathcal{S}$ is a data–driven stress scale defined from the median and median–absolute–deviation of the simulated stress signal, representing the smallest stress change the model can reliably resolve.
% More specifically, $\tau_{\mathcal{S}}$ is taken as the minimum stress difference that $M_{\mathcal{S}}$ can represent, computed from the model's parameter grid spacing together with a robust measure of the stress variation present in the data.
Differences smaller than $\tau_{\mathcal{S}}$ are not resolvable by $M_{\mathcal{S}}$ and therefore do not constitute genuine distinctions between the two models.
To ensure that $M_\mathcal{C}$ is penalized whenever it can emulate any lower-dimensional model, we take
\begin{equation}
    w_{\mathrm{red}}({\boldsymbol{\theta}}_{\mathcal{C}})=\min_{M_{\mathcal{S}} \subseteq M_{\mathcal{C}}}w_{\mathcal{C}\rightarrow\mathcal{S}}({\boldsymbol{\theta}}_{\mathcal{C}}).
\label{eq:wred}
\end{equation}

Finally, the prior for the parameter space is constructed by combining a harmonic-mean bottleneck of the form 
\begin{equation}
    H_i({\boldsymbol{\theta}}_i)=\frac{d}{\displaystyle\sum_{\ell=1}^{d}\frac{1}{{\boldsymbol{\theta}}_{i,\ell} + \varepsilon}},
\end{equation}
where $d$ is the model dimension and a small, machine precision value of $\varepsilon>0$ is included for numerical stability, and the redundancy penalty,
\begin{equation}
    P(\boldsymbol{\theta}_i|M_i)=H_i({\boldsymbol{\theta}}_i)w_{\mathrm{red}}({\boldsymbol{\theta}}_i)\label{eq:paramPrior},
\end{equation}
such that $\sum P(\boldsymbol{\theta}_i|M_i)=1$.
% With this, we obtain the likelihood $P(D|M_i)$.

Finally, we define the model prior.
Similar to the parameter prior, we consider the dimensionality of the models and their effect on the posterior.
We construct a weakly informative prior that penalizes model complexity,
\begin{equation}
    P(M_i) =\exp\left(-\frac{k_M}{2}\log N_{\text{eff}}\right),
\end{equation}
where $k_M$ is the number of free parameters in model $M_i$ and $N_{\text{eff}}$ is the effective number of scalar observations.
This prior is analogous to the Bayesian Information Criterion complexity term and introduces an explicit Occam penalty~\cite{Neath2012}.

\subsubsection{Posterior}
To compute the posterior $P(M_i|D)$, we  consider the marginal likelihood.
However, without additional prior information about the data, calculating the marginal likelihood is a challenging task.
Therefore, for the model comparison process, $P(D)$ is constant in the models to normalize the posteriors and, thus, the sum of the posteriors is unity~\cite{Gregory2005}:
\begin{equation*}
    \label{eq:normalization}
    \sum_{i=1}^{N_M}P(M_i|D)=1.
\end{equation*}

The likelihood and prior values can span orders of magnitude across grid points and trials, making direct summation numerically unstable. 
To maintain numerical precision and prevent underflow, multiplicative terms are computed in log space, and marginalizations are carried out using the log-sum-exp operation. 
Taking the logarithm of \cref{eq:BayesThm} yields
\begin{align*}
    \log{P(M_i| D)} 
    &= \log{\frac{P(M_i)P(D| M_i)}{P(D)}}\\
    &= \log{P(M_i)} + \log{P(D| M_i)} - \log{P(D)},
    \label{eq:logBayesThm}
\end{align*}
where $\log{P(M_i)}$ penalizes models according to prior information, $\log{P(D| M_i)}$ rewards models that fit the data well, and $\log{P(D)}$ serves as the normalization constant ensuring that posterior probabilities sum to unity. 
The effect of Occam's Razor becomes explicit in log space: the prior term $\log{P(M_i)} = -\tfrac{k_M}{2}\log N_{\text{eff}}$ penalizes models with greater complexity (i.e., larger parameter dimension $k_M=d$) relative to the effective number of observations $N_{\text{eff}}$.

\subsubsection{Additional quantities of interest}

After selecting the most plausible model $M^*$, we compute two additional posterior distributions of interest using the same likelihood evaluations and parameter grid employed during the computation of the model likelihood $P(D| M_i)$. 
The observational likelihood, quantifies the probability of observing the data under a given model after integrating out both the model parameters and the noise scale:
\begin{equation*}
    P(D| M_i) = \int\cdots\int P(D| M_i,\boldsymbol{\theta}_i,\beta)P(\boldsymbol{\theta}_i| M_i)P(\beta)d\boldsymbol{\theta}_i d\beta.
\end{equation*}
This integral is approximated using the same discrete grids in $\boldsymbol{\theta}_i$ and $\beta$ that were used for the likelihood evaluations, so no additional forward simulations are required.

The marginal posterior over parameters $\boldsymbol{\theta}_g$ is obtained by integrating out the noise scale $\beta$ using normalized prior weights over the discrete $\beta$ grid $\{\beta_{b}\}_{b=1}^{B}$~\cite{Gelman2021,Sivia2006}:
\begin{equation}
    P(\boldsymbol{\theta}_g|D,M_i)
    \approx \sum_{b=1}^{B} 
    w_b\, P(D | M_i, \boldsymbol{\theta}_g, \beta_{b}),
    \label{eq:theta_marginal}
\end{equation}
where the normalized prior weights $w_b$ approximate the prior mass over $\beta$ via
\begin{equation*}
    w_b \propto P(\beta_{b}),
    \qquad
    \sum_{b=1}^{B} w_b = 1.
\end{equation*}
These weights arise from discretizing the continuous marginalization integral over $\beta$ (\cref{eq:betamarginal}) into a quadrature sum, where each $w_b$ corresponds to the probability content of the local interval around $\beta_{b}$ under the prior $P(\beta)$. 
% This approach is used in Bayesian numerical integration when analytical marginalization is intractable.

The prior over $\boldsymbol{\theta}_g$ is uniform across grid points. 
The resulting discrete posterior distribution is normalized such that
\begin{equation*}
    \sum_g P(\boldsymbol{\theta}_g | D,M_i) = 1,
\end{equation*}
and the maximum a posteriori (MAP) parameter estimate, $\boldsymbol{\theta}_{\mathrm{MAP}}$, corresponds to the grid point $g$ maximizing \cref{eq:theta_marginal}.

The posterior over $\beta$ for a fixed parameter value $\boldsymbol{\theta}_g$ is given by
\begin{equation}
    P(\beta_{b} | D, M_i, \boldsymbol{\theta}_g) \propto P(D | M_i, \boldsymbol{\theta}_g, \beta_{b}) P(\beta_{b}),
    \label{eq:beta_posterior}
\end{equation}
which is normalized over $b$ so that $\sum_b P(\beta_{b} | D, M_i, \boldsymbol{\theta}_g) = 1$.
When $\boldsymbol{\theta}_g$ is chosen as $\boldsymbol{\theta}_{\text{MAP}}$, the MAP estimate $\beta_{\text{MAP}}$ is simply the $\beta_{b}$ that maximizes \cref{eq:beta_posterior}.
Maximizing the posterior over $\beta$ provides a principled estimate of the most likely noise scale in the data for the chosen model, which serves as a direct measure for the uncertainty quantification (UQ) of the fitted dynamics.
A larger $\beta_{\text{MAP}}$ represents greater experimental variability and correspondingly wider predictive intervals. 
In contrast, a smaller $\beta_{\text{MAP}}$ reflects higher confidence in the model's ability to reproduce the observed data within the inferred noise bounds.

\Cref{alg:bIMRalgorithm} shows the hierarchical Bayesian inference workflow, including heteroscedastic likelihood evaluation, marginalization over the noise scale parameter $\beta$, model likelihood computation, and extraction of $\boldsymbol{\theta}_{\mathrm{MAP}}$ and $\beta_{\mathrm{MAP}}$. 
The algorithm also outputs the posterior distributions $P(\beta | D,M_i)$ and $P(\boldsymbol{\theta}_i |D,M_i)$ for UQ.

\begin{algorithm}[t]
    \caption{Hierarchical Bayesian model selection for IMR}
    \label{alg:bIMRalgorithm}
    \begin{algorithmic}[1]
        \State \textbf{Input:} Experimental data $D = \{t^*_{j,k},R^*_{j,k},\dot{R}^*_{j,k}\}_{k\in\mathcal{K}_j}$; set of candidate models $M_i$ with parameter grid $\{\boldsymbol{\theta}_{i,g}\}$; uniform $\beta$ grid $\{\beta_{b}\}$ with prior $P(\beta)$
        \State Compute baseline variances $\sigma^2_{0,R^*}(t^*_{j,k})$, $\sigma^2_{0,\dot{R}^*}(t^*_{j,k})$; strain rates $\dot{\varepsilon}^*_{j,k}$, activity $a_{j,k}$ and weights $w_{j,k}$ 
    \State Using the stress integral mismatch measures $\Phi_{\mathcal{C}\rightarrow\mathcal{S}}$ and redundancy factors $w_{\mathcal{C}\rightarrow\mathcal{S}}$ [\cref{eq:PhiCtoS,eq:wCtoS}], compute the parameter–space priors $P(\boldsymbol{\theta}_i|M_i)$ for all models via \cref{eq:paramPrior}
    \For{each model $M_i$}
            \For{each grid point $\boldsymbol{\theta}_{i,g}$}
                \State Interpolate simulations to trial $j$'s $t^*_{j,k}$
                \For{each $\beta_{b}$}
                    \State Evaluate \cref{eq:obsLklhd} using Gaussian likelihood with heteroscedastic variance
                \EndFor
                \State Marginalize over $\beta$ to obtain \cref{eq:betamarginal}
            \EndFor
            \State Compute model likelihood, \cref{eq:theta_quadrature}, by summing over $\boldsymbol{\theta}_{i,g}$
        \EndFor
        \State Compute posterior model probabilities, \cref{eq:BayesThm}, and select best model $M^*$
        \State Optionally, compute: marginal parameter posterior, \cref{eq:theta_marginal}, and noise scale posterior, \cref{eq:beta_posterior}, to obtain $\boldsymbol{\theta}_{MAP}$ and $\beta_{\text{MAP}}$, respectively
        \State \textbf{Output:} $P(M_i | D)$, $P(\boldsymbol{\theta}_{i,g} | D,M_i)$, $P(\beta | D,M_i)$, $M^*$, $\boldsymbol{\theta}_{\mathrm{MAP}}$, $\beta_{\mathrm{MAP}}$
    \end{algorithmic}
\end{algorithm}

\section{Results\label{sec:results}}

\subsection{Consistency check}

% To validate our Bayesian inference approach, 
We perform a consistency check to recover the model and parameter values of synthetic data for which the true viscoelastic model and parameters are known a priori.
The consistent model selection from synthetic ground-truth conditions ensure high fidelity inverse characterization of the experimental data.
Using the bubble dynamics model described in \cref{subsec:physics} along with the stress integral formulations in \cref{tab:models}, $32$ sets of $R^*(t^*)$ synthetic data for each of the constitutive models are simulated.
The parameter values used to generate synthetic data were chosen for their relevance to the soft materials of interest.
Specifically, the values used were $\mu=\SI{0.05}{\pascal\second}$, $G=\SI{10}{\kilo\pascal}$, $\lambda_1=\SI{1e-5}{\second}$, and $\alpha=1$.
$R_{\text{max}}$ is varied with respect to the observed experimental LIC data from \cref{fig:lambda_vs_Rmax}.
Each constitutive model is then compared to the synthetic data using the approach in \cref{sec:characterize}.
The results are tabulated in \cref{tab:synthetic-fit}.
The method correctly identifies generative models that serve as the ground-truth with close parameter values and posterior values equal to unity.

% For the most plausible model, synthetic data generated with models with $\mu=0.1$ Pa s, the method recovered values in the range \SIrange{0.05}{0.11}{\pascal\second}; for $G=\SI{15}{\kilo\pascal}$ it recovered values in the range \SIrange{5.32}{16.6}{\kilo\pascal}; for $\lambda_1=10^{-5}$, it recovered values in the range $\SIrange{1.4e-5}{2.2e-5}{\second}$; for $\alpha=1$, it recovered the value 0.86.
% However, while the method correctly identified the true generating model for most cases, it was not always the most plausible.
% Specifically, for the qNH ground-truth data, the best-supported model was the three-parameter qKV model.
% This result is reasonable since the stress integral formulation for the qNH model is a subset of the qKV model.
% For data generated using the qNH model, the extracted values from the qKV model were $\mu=0.002$ Pa s, $G=16.6$ kPa, and $\alpha=0.86$.
% Thus, the likelihood gains obtained by allowing viscous dissipation in the qKV model outweigh the model complexity penalty, shifting the likelihood in favor of the qKV model.

\begin{table}[htbp]
\centering
\footnotesize
\caption{Consistency check using synthetic data with known parameter values. 
The rows highlighted in blue represent the most plausible model with a posterior value of unity.}
\label{tab:synthetic-fit}
\begin{tabular}{c|c|c|c|c|c}
\toprule
Ground-truth Model & Model & $\mu$ (\si{\pascal\second}) & $G$ (\si{\kilo\pascal}) & $\lambda_1$ (\si{\second}) & $\alpha$ \\
\midrule
& \cellcolor{blue!20}Newt & \cellcolor{blue!20}0.051 & \cellcolor{blue!20}-- & \cellcolor{blue!20}-- & \cellcolor{blue!20}--\\
& NH & -- & 10.6 & -- & --\\
Newt & KV & 0.046 & 0.58 & -- & --\\
$\mu=\qty{0.05}{\pascal\second}$ & qNH & -- & 9.91 & -- & 0.009\\
& LM & 0.072 & -- & \SI{1.0e-7}{} & --\\
& qKV & 0.046 & 0.55 & -- & 0.012\\
& SLS & 0.046 & 3.02 & \SI{1.0e-7}{} & --\\
\midrule
& Newt & 0.049 & -- & -- & --\\
& \cellcolor{blue!20}NH & \cellcolor{blue!20}-- & \cellcolor{blue!20}10.4 & \cellcolor{blue!20}-- & \cellcolor{blue!20}--\\
NH & KV & 0.001 & 9.91 & -- & --\\
$G=\qty{10}{\kilo\pascal}$ & qNH & -- & 9.91 & -- & 0.014\\
& LM & 0.062 & -- & \SI{1.0e-7}{} & --\\
& qKV & 0.007 & 3.02 & -- & 0.464\\
& SLS & 0.014 & 9.39 & \SI{1.85e-7}{} & --\\
\midrule
& Newt & 0.138 & -- & -- & --\\
& NH & -- & 20.4 & -- & --\\
KV & \cellcolor{blue!20}KV & \cellcolor{blue!20}0.054 & \cellcolor{blue!20}9.91 & \cellcolor{blue!20}-- & \cellcolor{blue!20}--\\
$\mu=\qty{0.05}{\pascal\second}$ & qNH & -- & 19.5 & -- & 0.001\\
$G=\qty{10}{\kilo\pascal}$ & LM & 0.173 & -- & \SI{1.79e-7}{} & --\\
& qKV & 0.046 & 9.39 & -- & 0.022\\
& SLS & 0.086 & 9.39 & \SI{1.0e-7}{} & --\\
\midrule
& Newt & 0.423 & -- & -- & --\\
& NH & -- & 88.0 & -- & --\\
qNH & KV & 0.0193 & 65.8 & -- & --\\
$G=\qty{10}{\kilo\pascal}$ & \cellcolor{blue!20}qNH & \cellcolor{blue!20}-- & \cellcolor{blue!20}13.0 & \cellcolor{blue!20}-- & \cellcolor{blue!20}0.720\\
$\alpha=1$ & LM & 0.268 & -- & \SI{8.96e-7}{} & --\\
& qKV & 0.001 & 16.6 & -- & 0.464\\
& SLS & 0.293 & 91.0 & \SI{1.17e-6}{} & --\\
\midrule
& Newt & 0.001 & -- & -- & --\\
& NH & -- & 5.54 & -- & --\\
LM & KV & 0.001 & 5.77 & -- & --\\
$\mu=\qty{0.05}{\pascal\second}$ & qNH & -- & 4.41 & -- & 0.093\\
$\lambda_1=\SI{1e-5}{\second}$ & \cellcolor{blue!20}LM & \cellcolor{blue!20}0.054 & \cellcolor{blue!20}-- & \cellcolor{blue!20}\SI{5.99e-6}{} & \cellcolor{blue!20}--\\
& qKV & 0.001 & 3.02 & -- & 0.251\\
& SLS & 0.086 & 1.71 & \SI{1.36e-5}{} & --\\
\midrule
& Newt & 0.423 & -- & -- & --\\
& NH & -- & 57.1 & -- & --\\
qKV & KV & 0.072 & 75.3 & -- & --\\
$\mu=\qty{0.05}{\pascal\second}$ & qNH & -- & 38.3 & -- & 0.125\\
$G=\qty{10}{\kilo\pascal}$ & LM & 0.359 & -- & \SI{2.78e-7}{} & --\\
$\alpha=1$ & \cellcolor{blue!20}qKV & \cellcolor{blue!20}0.046 & \cellcolor{blue!20}16.6 & \cellcolor{blue!20}-- & \cellcolor{blue!20}0.464\\
& SLS & 0.046 & 51.6 & \SI{1.0e-7}{} & --\\
\midrule
& Newt & 0.373 & -- & -- & --\\
& NH & -- & 2.56 & -- & --\\
SLS & KV & 0.481 & 29.2 & -- & --\\
$\mu=\qty{0.05}{\pascal\second}$ & qNH & -- & 7.57 & -- & 1.49\\
$G=\qty{10}{\kilo\pascal}$ & LM & 0.359 & -- & \SI{1.0e-7}{} & --\\
$\lambda_1=\SI{1e-5}{\second}$ & qKV & 0.541 & 29.2 & -- & 0.002\\
& \cellcolor{blue!20}SLS & \cellcolor{blue!20}0.293 & \cellcolor{blue!20}29.2 & \cellcolor{blue!20}\SI{1.36e-5}{} & \cellcolor{blue!20}--\\
\bottomrule
\end{tabular}
\end{table}

\begin{table*}[htpb]
\centering
\caption{Plausibilities for UM and UT experimental data and corresponding parameter values for each model. 
The rows highlighted in blue represent the most plausible model.}
\label{tab:expdata-fit}
\begin{tabular}{c|c|c|c|c|c}
\toprule
Material & Model &$\mu$ (\si{\pascal\second}) & $G$ (\si{\kilo\pascal}) & $\lambda_1$ (\si{\second}) & $\alpha$\\
\midrule
& Newt & 0.058 & -- & -- & -- \\
& NH & -- & 20.2 & -- & -- \\
& \cellcolor{blue!20}KV & \cellcolor{blue!20}0.022 & \cellcolor{blue!20}13.0 & \cellcolor{blue!20}-- & \cellcolor{blue!20}-- \\
UM1 & qNH & -- & 19.5 & -- & 0.01 \\
& LM & 0.083 & -- & \SI{1.16e-7}{} & -- \\
& qKV & 0.025 & 9.39 & -- & 0.074 \\
& SLS & 0.046 & 16.6 & \SI{1.85e-7}{} & -- \\
\midrule
& Newt & 0.148 & -- & -- & -- \\
& NH & -- & 7.71 & -- & -- \\
& \cellcolor{blue!20}KV & \cellcolor{blue!20}0.173 & \cellcolor{blue!20}9.91 & \cellcolor{blue!20}-- & \cellcolor{blue!20}-- \\
UM2 & qNH & -- & 7.57 & -- & 0.003 \\
& LM & 0.173 & -- & \SI{1.0e-7}{} & -- \\
& qKV & 0.158 & 9.39 & -- & 0.003 \\
& SLS & 0.001 & 1.71 & \SI{4.64e-5}{} & -- \\
\midrule
& Newt & 0.195 & -- & -- & -- \\
& NH & -- & 41.6 & -- & -- \\
& \cellcolor{blue!20}KV & \cellcolor{blue!20}0.112 & \cellcolor{blue!20}22.3 & \cellcolor{blue!20}-- & \cellcolor{blue!20}-- \\
UM3 & qNH & -- & 38.3 & -- & 0.007 \\
& LM & 0.173 & -- & \SI{1.16e-7}{} & -- \\
& qKV & 0.158 & 5.32 & -- & 0.464 \\
& SLS & 0.086 & 29.2 & \SI{1.0e-7}{} & -- \\
\midrule
& Newt & 0.125 & -- & -- & -- \\
& NH & -- & 45.2 & -- & -- \\
& \cellcolor{blue!20}KV & \cellcolor{blue!20}0.072 & \cellcolor{blue!20}33.5 & \cellcolor{blue!20}-- & \cellcolor{blue!20}-- \\
UT1 & qNH & -- & 43.9 & -- & 0.012 \\
& LM & 0.645 & -- & \SI{6.94e-6}{} & -- \\
& qKV & 0.086 & 29.2 & -- & 0.074 \\
& SLS & 0.086 & 29.2 & \SI{1.0e-7}{} & -- \\
\midrule
& Newt & 0.441 & -- & -- & -- \\
& NH & -- & 166.7 & -- & -- \\
& KV & 0.645 & 381.5 & -- & -- \\
UT2 & qNH & -- & 13.0 & -- & 4.81 \\
& LM & 0.416 & -- & \SI{2.40e-7}{} & -- \\
& \cellcolor{blue!20}qKV & \cellcolor{blue!20}0.158 & \cellcolor{blue!20}9.39 & \cellcolor{blue!20}-- & \cellcolor{blue!20}10.0 \\
& SLS & 1.00 & 500 & \SI{3.41e-7}{} & -- \\
\bottomrule
\end{tabular}
\end{table*}

\subsection{Inverse characterization via Bayesian model selection}
The Bayesian inference method was applied to the experimental UM and UT LIC datasets. 
The model plausibilities and MAP parameter estimates are tabulated in \cref{tab:expdata-fit}.
The optimal noise scaling parameter is included for the experimental data results to account for the variability of the experimental data.

\begin{figure}[t!]
  \centering
  \begin{subfigure}[t]{0.32\textwidth}
    \centering
    \includegraphics[width=\textwidth]{{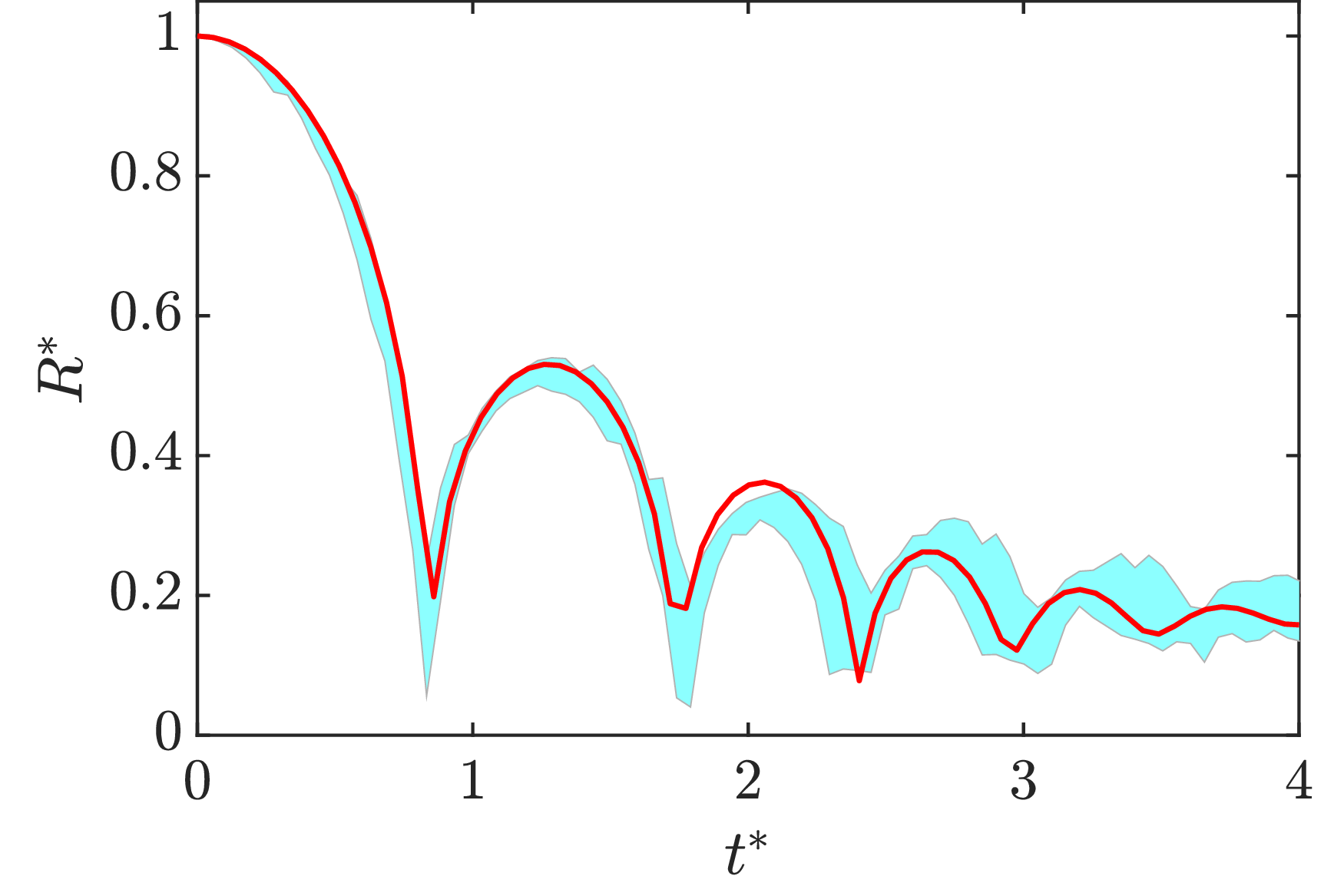}}
    \caption{UM1}
  \end{subfigure}\hfill
  \begin{subfigure}[t]{0.32\textwidth}
    \centering
    \includegraphics[width=\textwidth]{{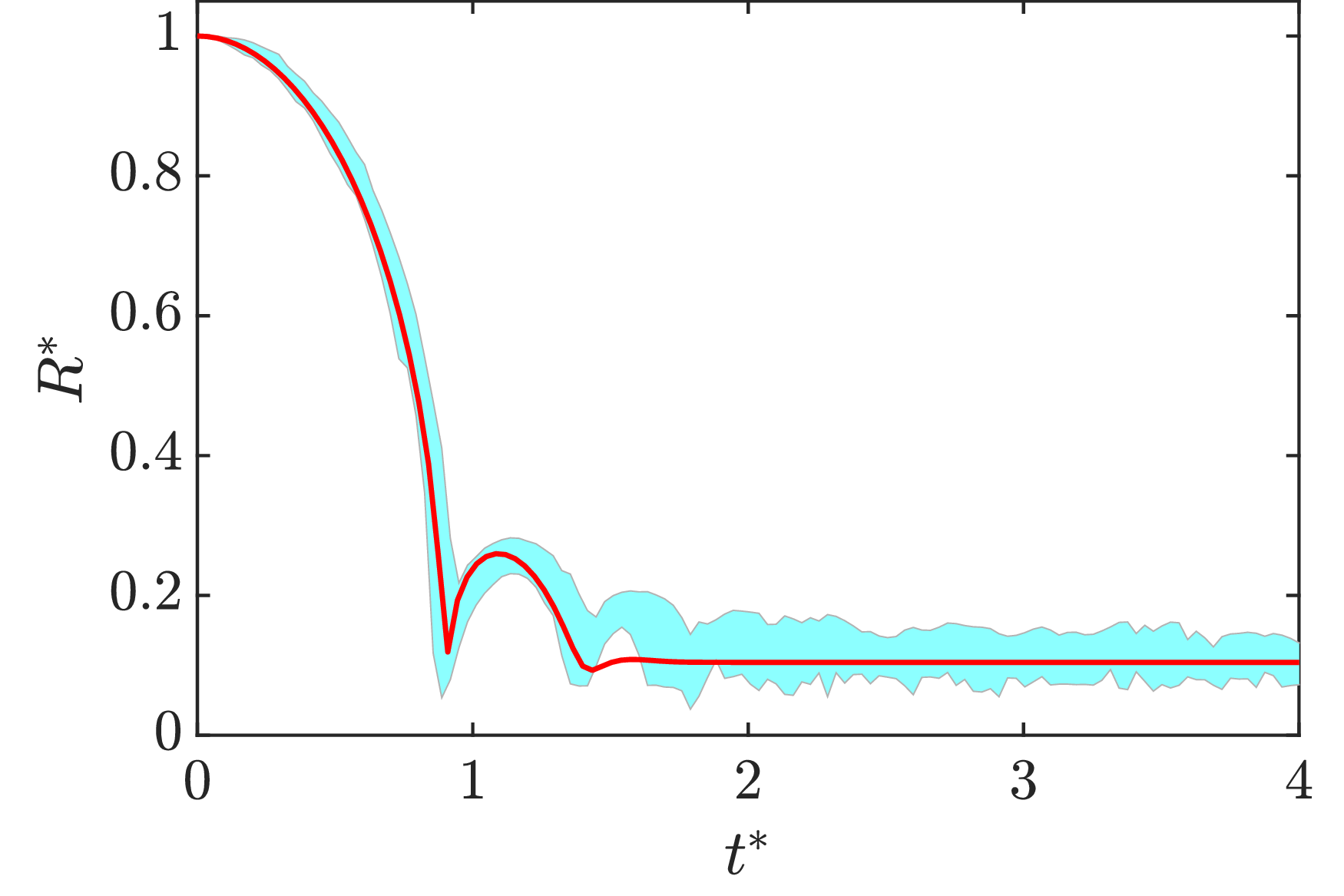}}
    \caption{UM2}
  \end{subfigure}\hfill
  \begin{subfigure}[t]{0.32\textwidth}
    \centering
    \includegraphics[width=\textwidth]{{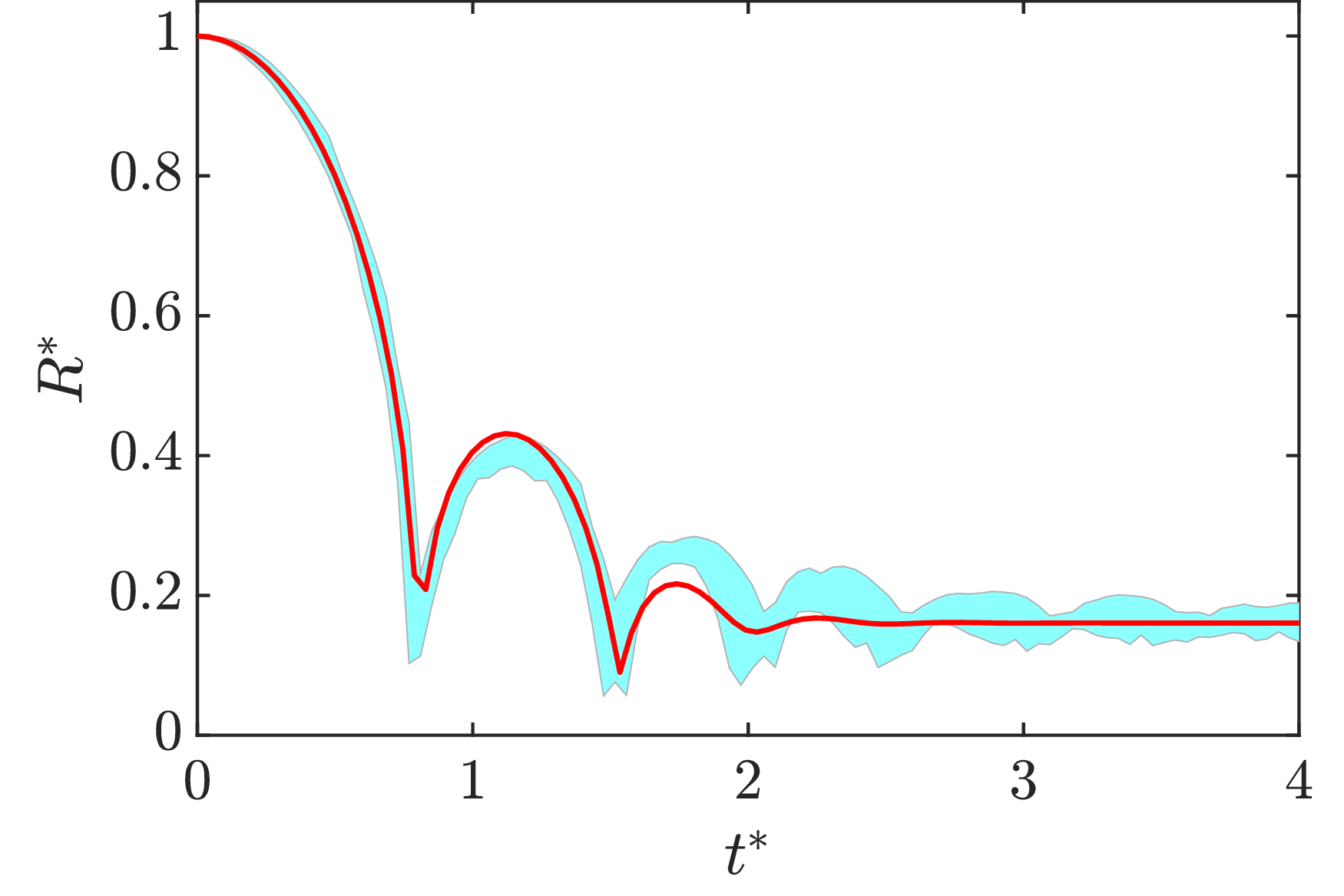}}
    \caption{UM3}
  \end{subfigure}

  \vspace{0.8em}

  \hspace*{0.12\textwidth}
  \begin{subfigure}[t]{0.32\textwidth}
    \centering
    \includegraphics[width=\textwidth]{{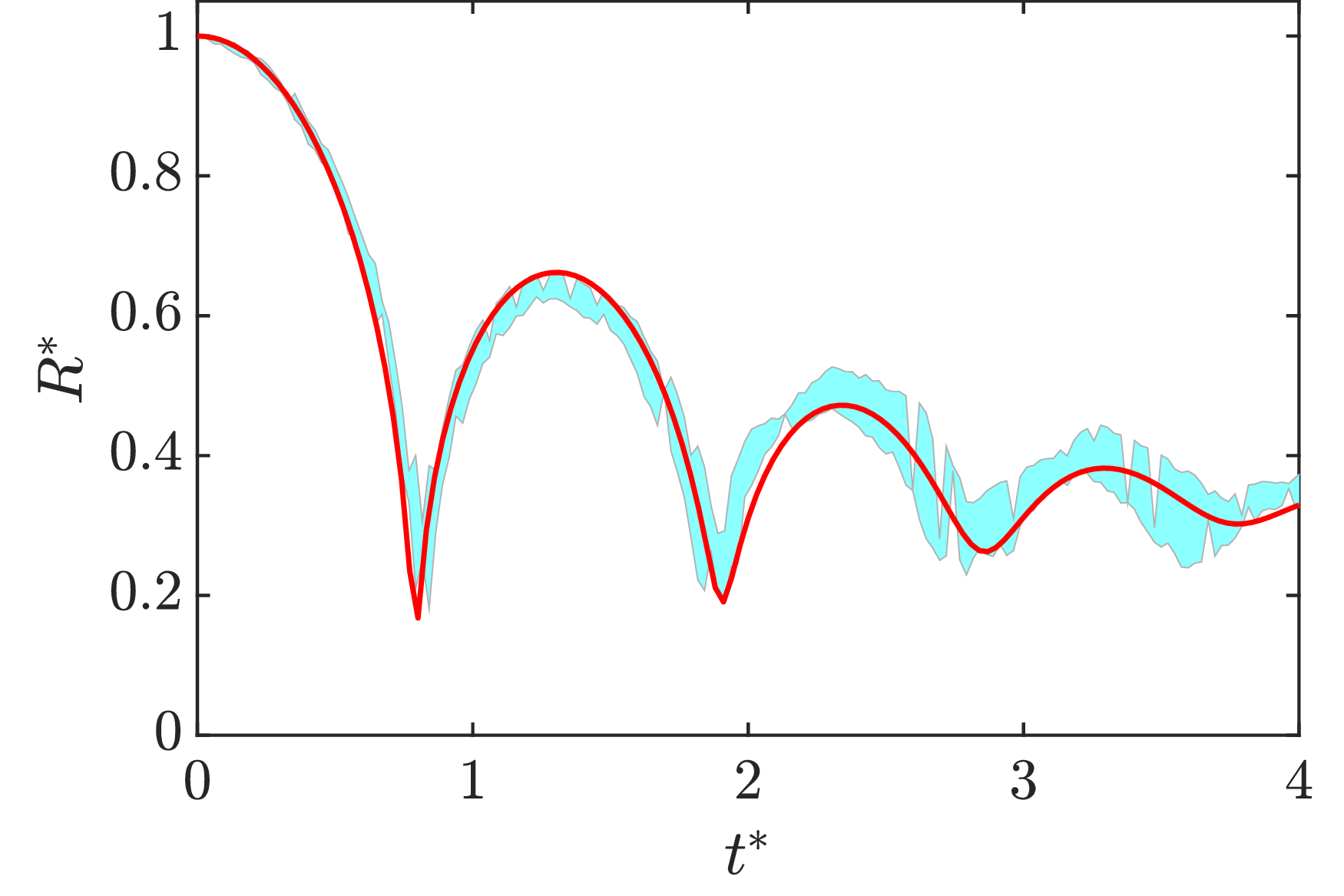}}
    \caption{UT1}
  \end{subfigure}
  \hspace*{0.04\textwidth}
  \begin{subfigure}[t]{0.32\textwidth}
    \centering
    \includegraphics[width=\textwidth]{{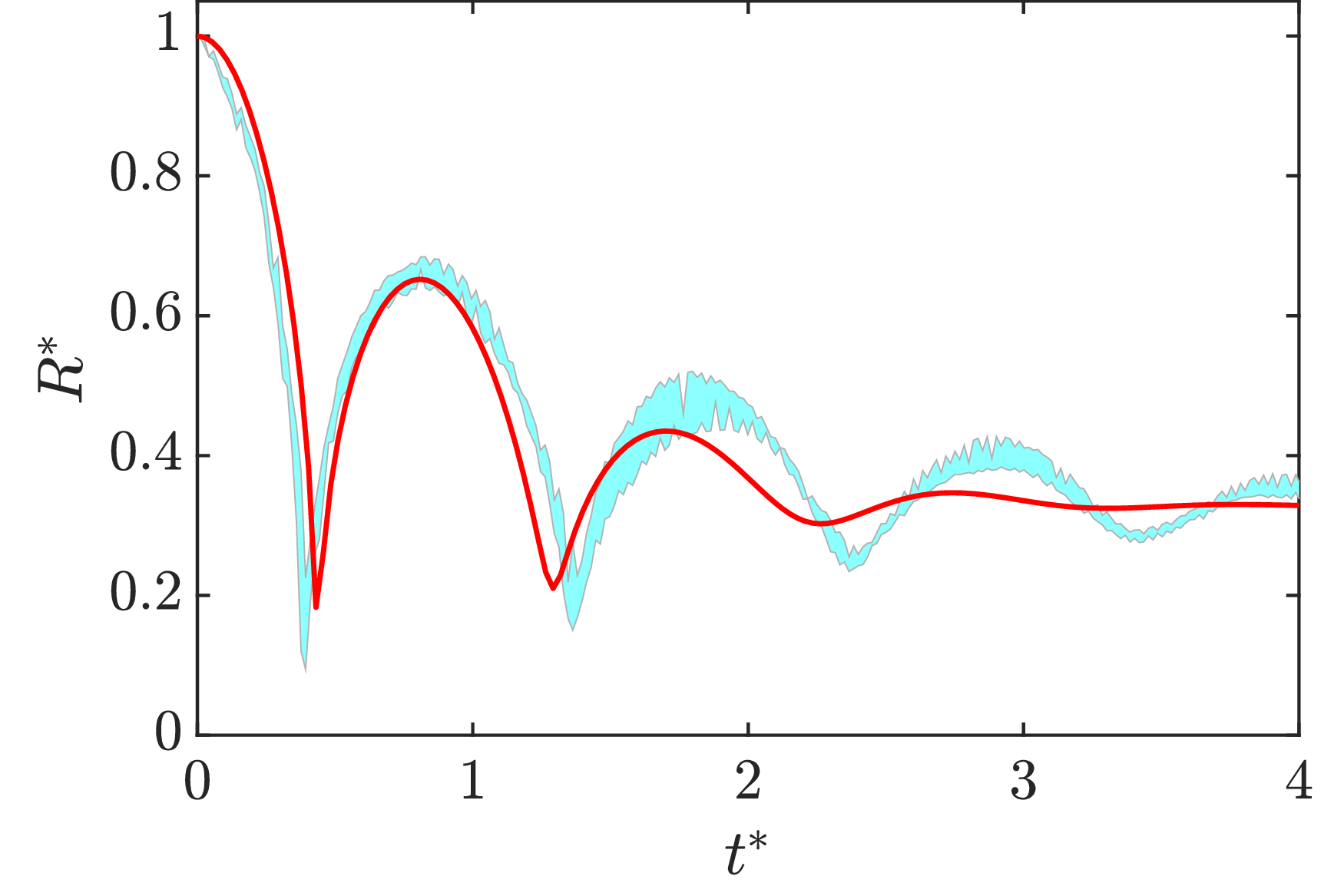}}
    \caption{UT2}
  \end{subfigure}
  \hspace*{0.16\textwidth}
  
  \caption{Normalized bubble radius experimental cloud $R^*$ versus non-dimensional time $t^*$. 
  Cyan region: experimental cloud across trials, showing variability in collapse and rebound dynamics. 
  Solid red curve: the simulation corresponding to the MAP parameter set $\boldsymbol{\theta}_{\text{MAP}}$ for the most plausible models(see  \cref{tab:expdata-fit}).}
  \label{fig:bestfits_all}
\end{figure}

Most datasets consistently placed the posterior mass on the KV model, with a single dataset favoring the qKV model.
For UM1, KV had a posterior of $0.83$ and qKV had a posterior of $0.17$.
For the other cases, the highlighted model in \cref{tab:expdata-fit} had a posterior of unity.
The finding that multiple materials select the same constitutive family does not imply identical behavior. 
Their MAP parameter estimates occupy distinct and well-separated regions of parameter space, reflecting genuine material differences.
In particular, the KV-selected materials differed primarily in their relative elastic-to-viscous values, shifting collapse-time sensitivity and rebound curvature in characteristic ways, as shown in \cref{fig:bestfits_all}.
The dataset selecting qKV showed a sharper rebound curvature and steeper post-collapse trajectory compared to the others, which the linear KV model could not reproduce.
The Bayesian likelihood correctly identified this mismatch and shifted posterior mass toward the nonlinear model.

\begin{figure}[!ht]
  \centering

  \begin{subfigure}[t]{0.32\textwidth}
    \centering
    \includegraphics[width=\textwidth]{{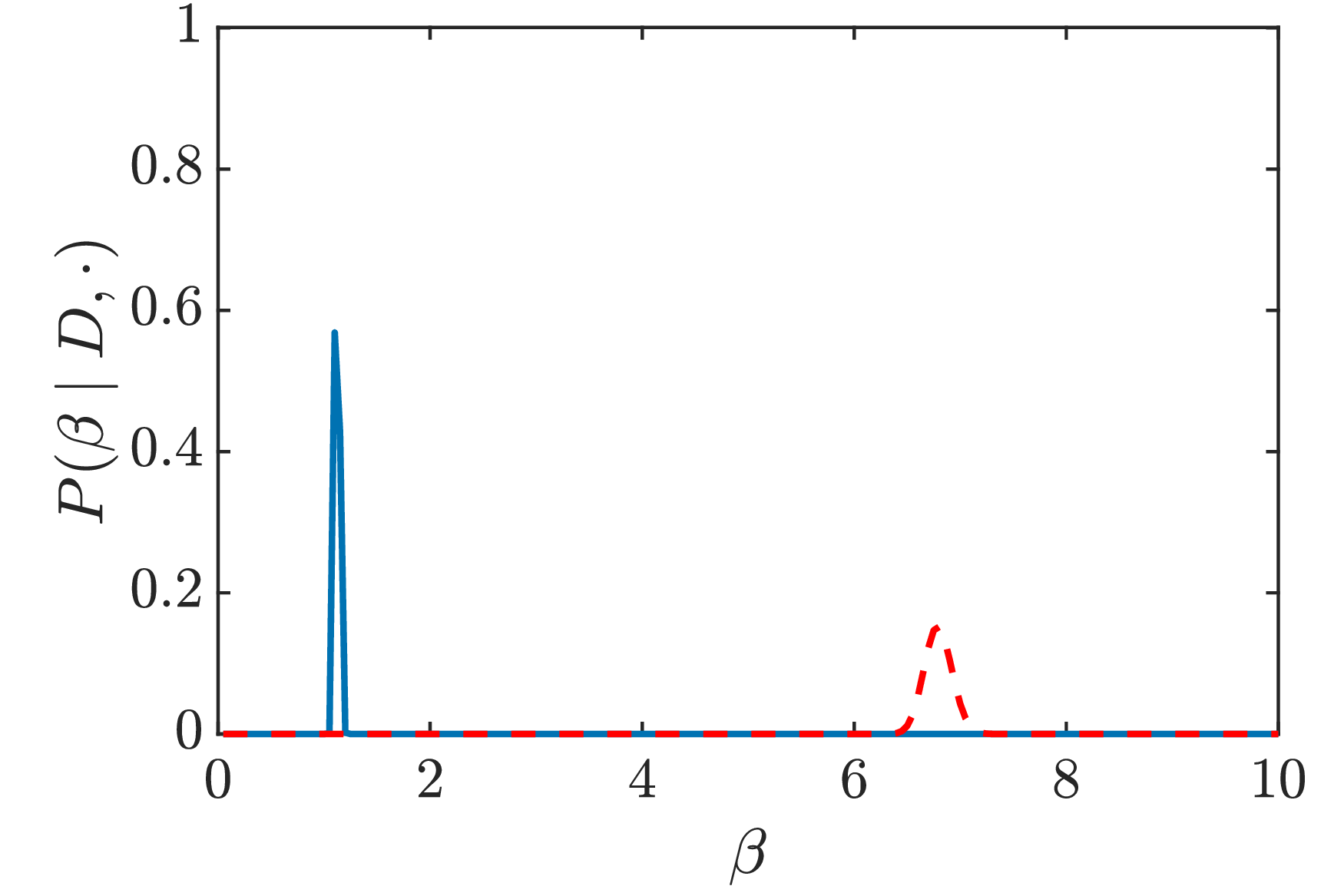}}
    \caption{UM1 ($M^*$ = KV)}
  \end{subfigure}\hfill
  \begin{subfigure}[t]{0.32\textwidth}
    \centering
    \includegraphics[width=\textwidth]{{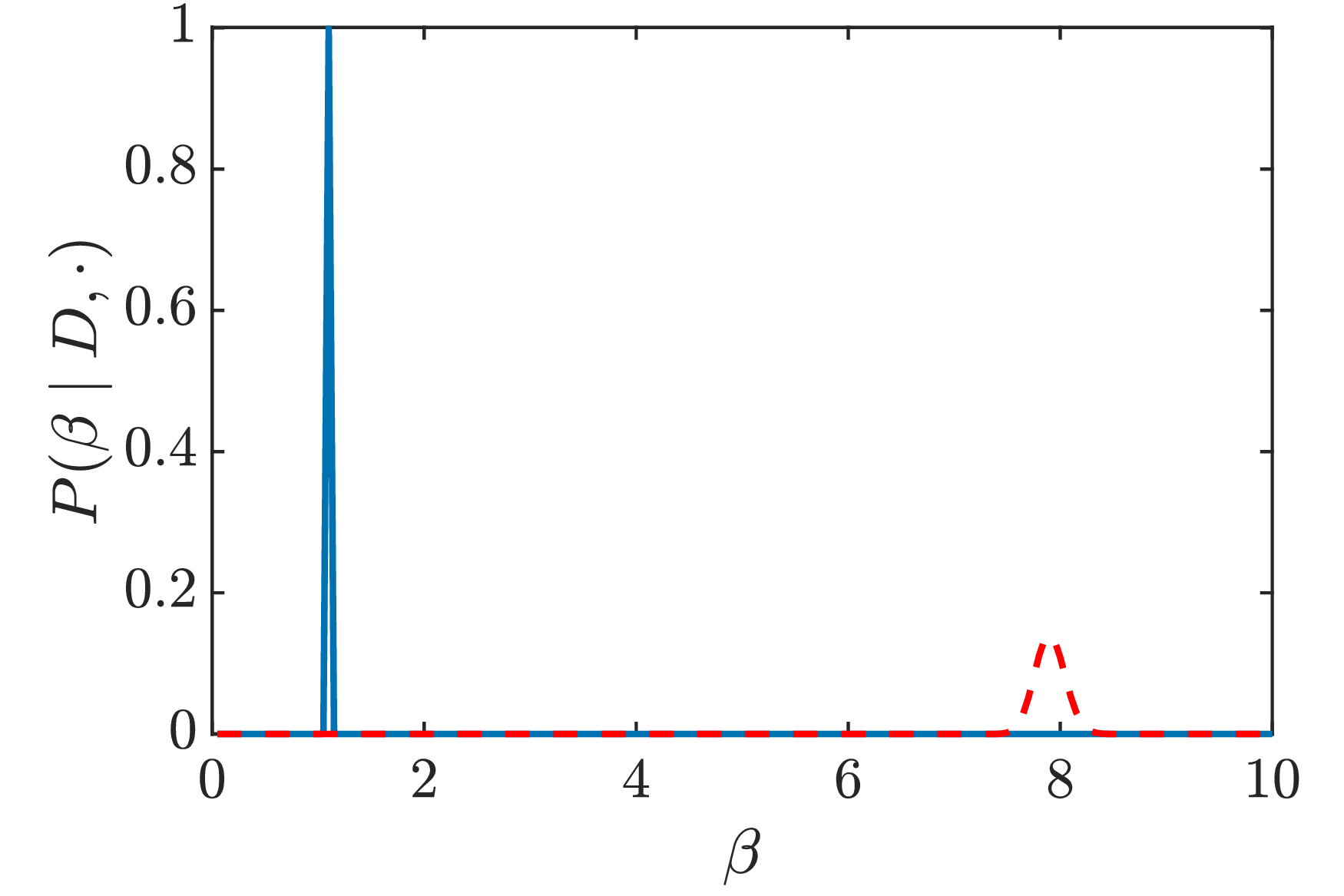}}
    \caption{UM2 ($M^*$ = KV)}
  \end{subfigure}\hfill
  \begin{subfigure}[t]{0.32\textwidth}
    \centering
    \includegraphics[width=\textwidth]{{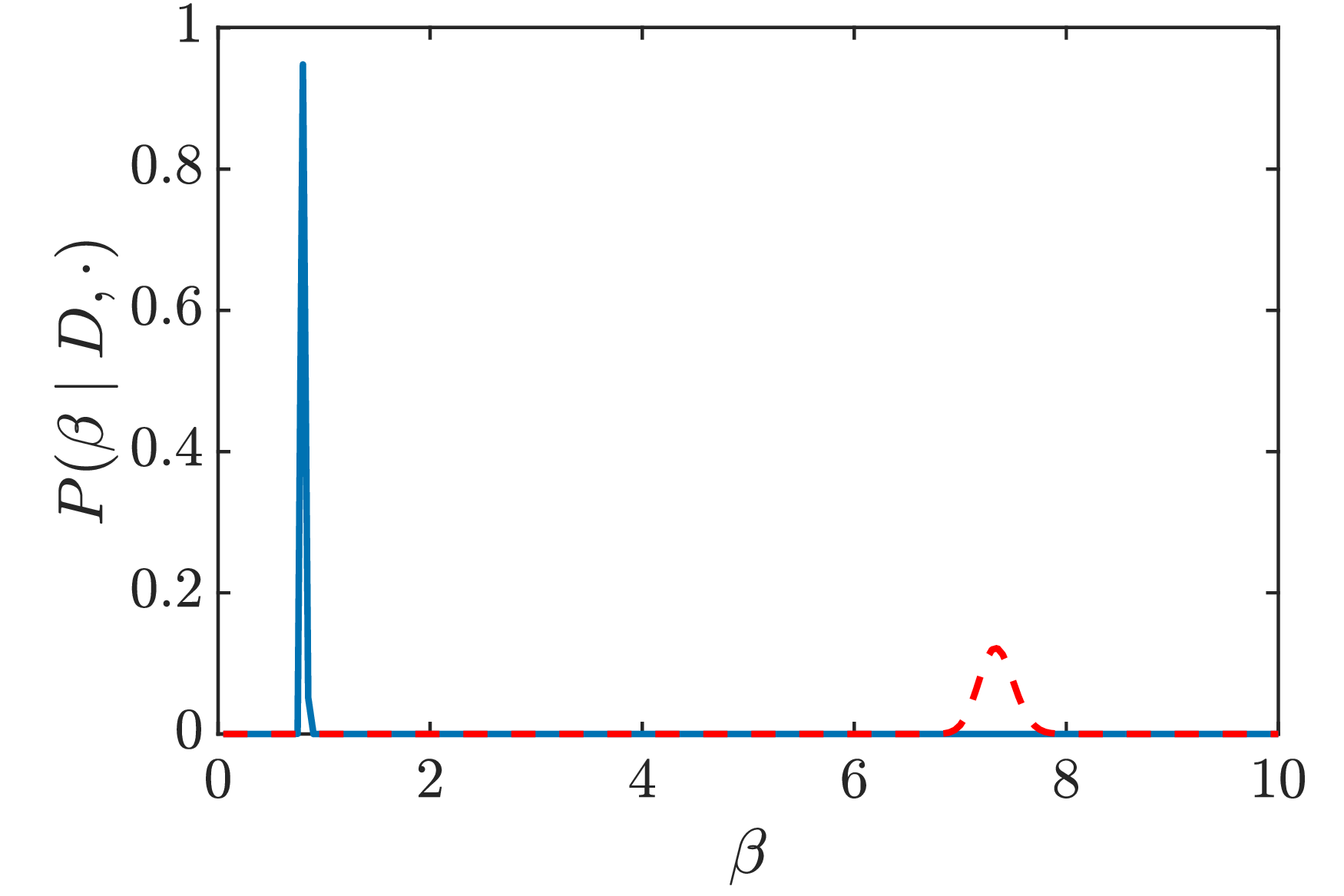}}
    \caption{UM3 ($M^*$ = KV)}
  \end{subfigure}

  \vspace{0.8em}

  \hspace*{0.12\textwidth}
  \begin{subfigure}[t]{0.32\textwidth}
    \centering
    \includegraphics[width=\textwidth]{{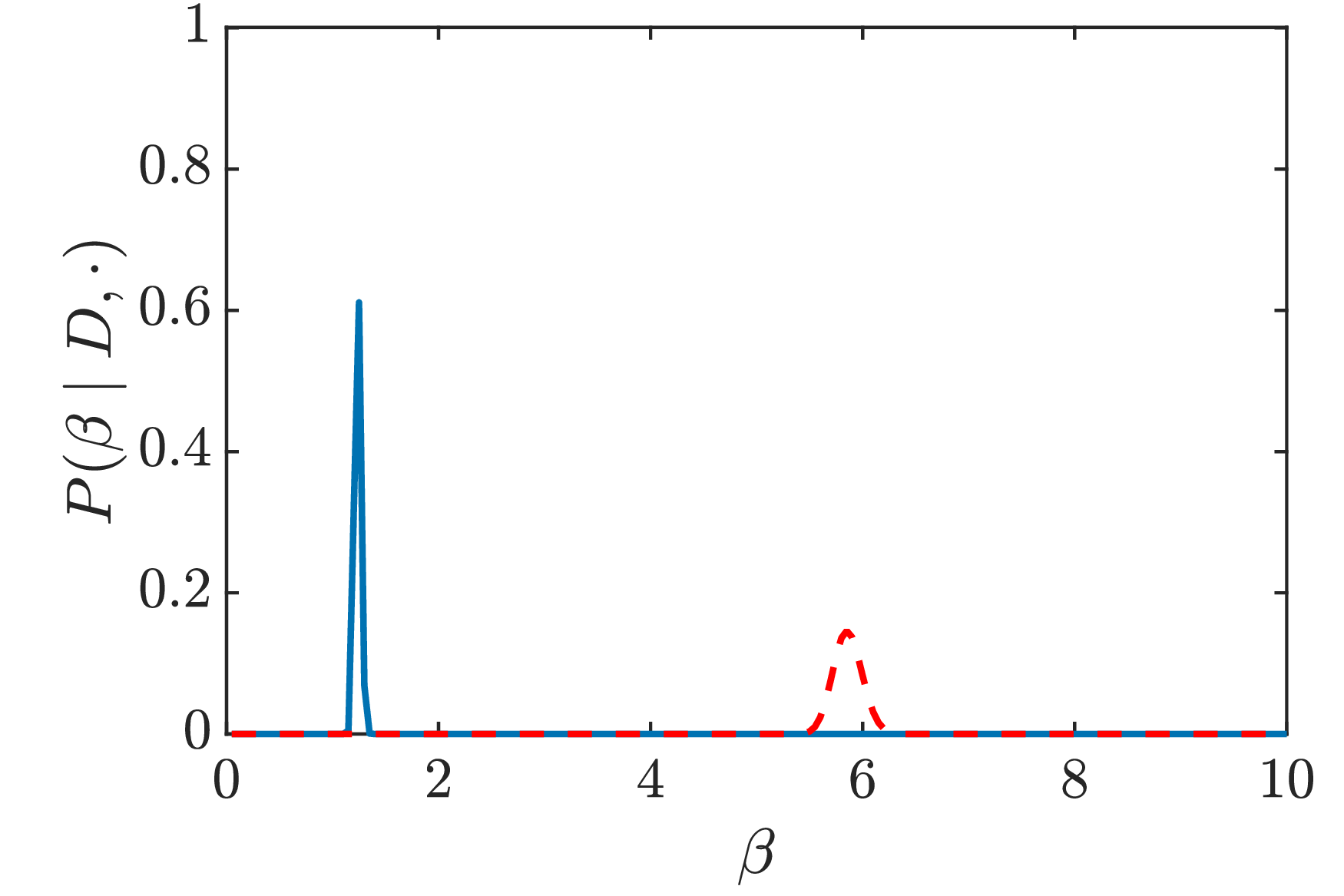}}
    \caption{UT1 ($M^*$ = KV)}
  \end{subfigure}
  \hspace*{0.04\textwidth}
  \begin{subfigure}[t]{0.32\textwidth}
    \centering
    \includegraphics[width=\textwidth]{{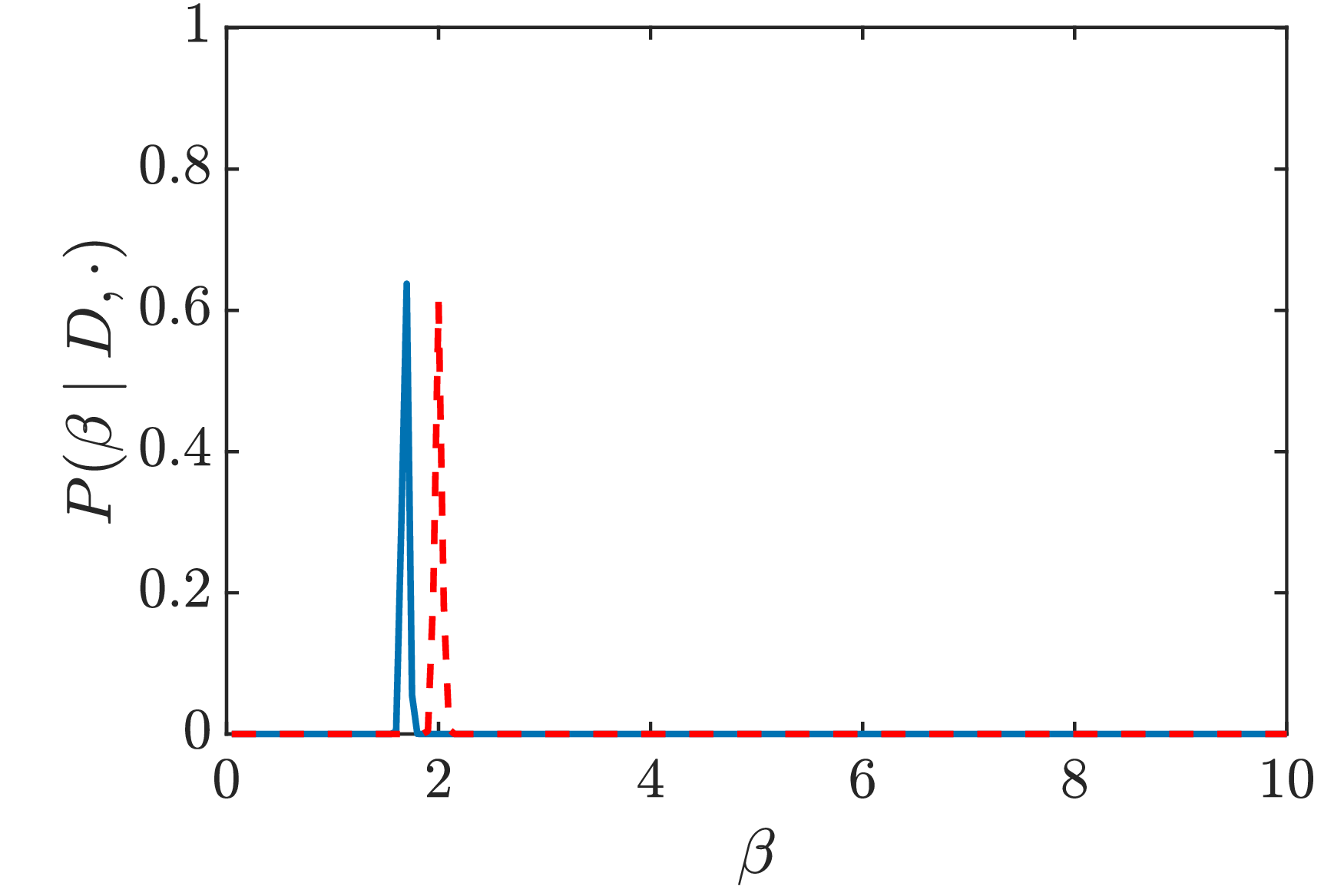}}
    \caption{UT2 ($M^*$ = qKV)}
  \end{subfigure}
  \hspace*{0.16\textwidth}

  \caption{Posterior uncertainty quantification for the noise scaling parameter $\beta$ in the most plausible model $M^*$. 
  Solid blue line: the marginal posterior $P(\beta|D,M^*)$ integrated over material parameters $\boldsymbol{\theta}$. 
  Dashed red line: the conditional posterior $P(\beta|D,M^*,\boldsymbol{\theta}_{\text{MAP}})$ evaluated at the MAP parameter set.}
  \label{fig:beta_uq}
\end{figure}

Finally, we examined the inferred noise-scale $\beta$ posteriors across materials.
Rather than acting as a free parameter that compensates for model mismatch, $\beta$ reflects how consistently the experimental trajectories align with the model-predicted collapse-rebound dynamics.
Datasets with trials following similar radius histories produce a sharply concentrated posterior for $\beta$, indicating that only a modest amount of measurement noise is required to explain the observed variation.
In contrast, datasets with visibly greater scatter yield broader noise-scale posteriors, reflecting genuine variability in the recorded bubble motion.
\Cref{fig:beta_uq} compares the marginal noise-scale posterior (solid blue), $P(\beta|D,M^*)$, with the conditional posterior at the MAP parameters (dashed red), $P(\beta|D,M^*,\boldsymbol{\theta}_{\text{MAP}})$. 
With the exception of the UT2 dataset, the red curves are broader and shifted toward larger $\beta$ than the blue curves, which are narrower and lower in value. %indicating that the best-fit parameter set alone inflates the effective noise level to accommodate localized mismatches.

\section{Discussion\label{sec:discussion}}
\subsection{Synthetic data insights}

The synthetic datasets provide a controlled environment for evaluating the Bayesian IMR method under known ground truths, showing both the strengths and limitations of the technique.
Across the synthetic cases, the method correctly recovered the true generating model and appropriate parameter estimates, but only when the informed priors were used. 
Under uniform priors, the model likelihood is dominated by the volume of the parameter space rather than the quality of fit.
In several synthetic tests, this led to incorrect model selection and favored higher-dimensional models (data not shown).
The constructed priors resolve this issue by weighting the parameter space according to the relative information. 
Regions that mimic simpler models receive reduced prior mass, preventing inflated support for unnecessarily complex models.
Specifically, the Bayesian likelihood penalizes regions of the parameter space unless they produce a genuine improvement in the dynamics.
The synthetic tests thus confirm that the priors are not an optional regularizer but an essential component that aligns the inference with the underlying physics.

Another consideration is the effect of the parameter-grid resolution on the computed likelihoods.
Since the likelihood is integrated across the entire grid, models with more parameters require exponentially more grid points to cover their domain with comparable resolution.
If the models use the same course grid (e.g., 16 values per dimension), the higher-dimensional models incorrectly dominate (data not shown).
This occurs because course grids over-smooth the likelihood landscape of the simpler models while still providing sufficient coverage for the complex ones, artificially inflating their likelihood. 
% Although these resolution-equalized tests are not shown, they consistently favored the most complex models, even when the data were generated by simpler ones.
For this reason, we use model-dependent, computationally tractable grids.

\subsection{Experimental findings}

The characterization of the experimental datasets reveals the strengths of the Bayesian IMR technique.
Overall, the selected model and parameters matched the early time collapse and rebound envelopes of the experimental clouds, with most materials placing their posterior mass on the KV model.
For the cross-institutional gelatin study, the Bayesian IMR technique identified the KV model as well for similar experimental setups and data acquisition methods.
The widespread selection of KV is due to its simplicity and accuracy as additional parameters from the other considered models did not provide any improvement in likelihood.
The only exception is the UT2 dataset, where the posterior favors the qKV model.
This shift arises from a reproducible feature in the first rebound where the data shows a much sharper recoil than the other materials.
The KV model cannot generate this feature without distorting the bubble radius history after the second collapse.
The nonlinear strain-stiffening term correctly adjusts the rebound slope without compromising the collapse timing.
Overall, these results confirm that the technique extracts constitutive behavior that is physically interpretable under realistic variability.

\subsection{UQ interpretability}

The posteriors for the noise scale parameter $\beta$ provide an interpretable measure of how confidently each constitutive model explains the experimental data. 
The marginal noise scale distribution, $P(\beta|D,M^*)$, summarizes the effective measurement uncertainty after integrating over plausible parameter combinations.
The conditional posterior at the MAP parameter values, $P(\beta|D,M^*,\boldsymbol{\theta}_{\text{MAP}})$, reflects the noise level required when the model is constrained to a single best-fit point.
When the red (conditional) curve is noticeably broader or shifted relative to the blue (marginal) curve, it indicates that the MAP parameter set alone absorbs localized mismatches between the model and the data. 
After marginalizing over nearby parameter values, those mismatches are partially accommodated by variations in $\boldsymbol{\theta}$, producing the narrower blue distributions.

This Bayesian formulation separates variability due to measurement noise from uncertainty in the parameters and from genuine model–data mismatch, making each source of error interpretable.
The noise parameter does not compensate for missing physics; it grows only in response to true measurement variability, not model-form error. 
As a result, the noise scale posteriors provide a transparent measure of confidence in the inferred material response and clearly distinguish between three sources of variability: experimental noise, parameter uncertainty, and model-form error.

% Thus, the behavior of the noise scale posteriors reflects whether model-data mismatch is dominated by experimental noise or model error.
When the posteriors coincide, the model performs consistently over the parameter space, whereas when the posteriors are separate, there is a clear parameter set that performs the best.
This coinciding behavior does not imply that only one dataset, e.g., UT2 under qKV, is well explained by its selected model.
Rather, the contraction or expansion of the noise scale posterior is dataset-specific. 
For UM1 and UT1, the marginal posterior remains sharply concentrated while the conditional posterior shifts to higher values and is broader.
This represents a case where the model can explain the data globally, yet localized misfits at the MAP parameters introduce small, structured residuals.

\subsection{Future directions}

While the Bayesian IMR method provides a rigorous platform for simultaneous model selection, parameter inference, and quantification of data uncertainty, several extensions remain open for exploration.
First, the current implementation relies on strain-based weighting, which prioritizes data points that meet the specified thresholds.
Future work could employ adaptive or data-driven weighting strategies, such as variational Bayesian or optimization-based strategies~\cite{Bishop2006,Rezende2014}, to dynamically balance sensitivity across the different strain-based regimes.
Second, although the hierarchical likelihood marginalizes over the noise scale $\beta$, the residuals are presently assumed to be Gaussian.
Relaxing assumptions through heavy-tailed or mixture models could improve robustness to experimental artifacts and outliers~\cite{Box2011,Sivia2006}.
Third, higher-dimensional parameter spaces may benefit from more efficient sampling methods, such as Markov Chain Monte Carlo, nested sampling, or surrogate accelerated inference~\cite{Sarkar2012,Gelman2021}.
Additionally, in cases where there may not be a clear selected model, Bayesian model averaging can propagate uncertainty in model choice rather than enforcing a single selected model, while cross-validation strategies can assess generalizability across datasets~\cite{Hoeting1999,Vehtari2017}.
Finally, integrating multi-modal experimental observables, such as digital image correlation fields or bubble asphericity \cite{mcghee2023high,yang2025inertial}, within a joint Bayesian likelihood could strengthen inference in cases where radius-time data alone are ambiguous.

Although the present study adopts the spherical stress integral formulation defined in~\cref{eq:stress_integral}, the Bayesian framework itself does not depend on this reduced representation.
Since the technique is independent of the forward solver, thus it could be extended to consider other constitutive material models.
More broadly, this formulation is a transferable method for Bayesian model selection and uncertainty quantification in nonlinear, data-limited systems where understanding rate-dependent constitutive behavior is critical.
% For example, in other continuum mechanics settings, the Cauchy stress tensor $\boldsymbol{\sigma}(\bold{\text{x}},t)$ could be computed by a different forward model, e.g., finite element simulations, and the likelihood constructed from experimentally measurable quantities.
Thus, the stress integral formulation used in IMR is a case of a broader class of Bayesian model selection problems in computational physics.
Other potential applications include the characterization of biological tissues and soft organs under impact or ultrasound stimulation, as well as cavitation- or fracture-induced failure in elastomers and soft composites.
Additionally, this method can be used to study high-rate compaction or damage evolution in geological and structural systems.
Since the approach integrates physics-informed forward modeling with hierarchical noise treatment and likelihood-based model selection, it can quantify both parametric and structural uncertainty across different parameter spaces.

\section{Conclusions\label{sec:conclusions}}

We formulated a hierarchical Bayesian method for inertial microcavitation rheometry of soft hydrogels, combining large-scale forward simulation libraries with statistically grounded model selection.
By constructing physically informed priors when evaluating the likelihood across a grid of constitutive models and parameter values, the method discriminates among competing models while providing \textit{maximum a posteriori} parameter estimates that serve as informed initial guesses. 
Synthetic data tests show that the method recovers the generating models and parameters and subsets of more complex models can lead to model misidentification if uninformed priors are used.
The KV model consistently described the UM datasets and UT1, while UT2 needed the nonlinear stiffening term in the qKV model.
The selected models reproduced the collapse and rebound windows, but in some cases, agreement was observed for the first few collapses and subsequently averaged out later noisy oscillations.
A comparison between gelatin samples of the same concentration from the University of Michigan and the University of Texas at Austin revealed that the presented method identifies a consistent constitutive model. 
Models that rely on artificially large noise parameter $\beta$ values are automatically downweighted in their likelihood, ensuring that plausibility reflects both fit quality and experimental variability.
Balancing between simplicity and flexibility, the method offers a ranking of constitutive models, interpretable parameter estimates, and uncertainty diagnostics.
By unifying constitutive inference, model selection, and uncertainty quantification, the Bayesian IMR method provides a transferable, probabilistic method for dynamic rheometry of soft matter and polymer materials.

\section{Acknowledgments}

VS and MRJ acknowledge Dr.~Tianyi~Chu for the useful conversations in the preparation of this manuscript.
MRJ acknowledges support from the U.S. Department of Defense under the DEPSCoR program Award No. FA9550-23-1-0485 (PM Dr.~Timothy~Bentley).
JBE, MRJ, and JY gratefully acknowledge support from the U.S. National Science Foundation (NSF) under Grant Nos. 2232426, 2232427, and 2232428, respectively.
JBE and SHB acknowledge support from the U.S. Department of Defense, the Army Research Office under Grant No. W911NF-23-10324 (PMs Drs.\ Denise~Ford and Robert~Martin).
This work used Anvil at Purdue University through allocation MCH220010 from the Advanced Cyberinfrastructure Coordination Ecosystem: Services \& Support (ACCESS) program, which is supported by U.S. National Science Foundation grants \#2138259, \#2138286, \#2138307, \#2137603, and \#2138296.
The opinions, findings, and conclusions, or recommendations expressed are those of the authors and do not necessarily reflect the views of the funding agencies.

\bibliographystyle{elsarticle-num-names}
\bibliography{ref}

\end{document}